\documentclass[a4paper,fleqn,usenatbib,useAMS]{mnras}
\usepackage{graphicx}	
\usepackage{amsmath}	
\usepackage{amssymb}    
\usepackage{amssymb}	
\usepackage{multicol}	
\usepackage{bm}		
\usepackage{pdflscape}	
\usepackage{hyperref}
\usepackage{lineno}

\newcommand{\redmapper}{redMaPPer}

\newcommand{\de}{\text{d}}

\newcommand{\LCDM}{$\Lambda$CDM }

\newcommand{\ob}{^{\rm ob}}
\newcommand{\true}{^{\rm true}}
\usepackage[usenames]{color}
\definecolor{purple}{RGB}{150,0,200}

\usepackage[T1]{fontenc}
\usepackage{ae,aecompl}

\usepackage{txfonts}

\title[Modeling projection effects in optically-selected cluster catalogues]{Modeling projection effects in optically-selected cluster catalogues}

\author[DES Collaboration]{M.~Costanzi$^{1}$\thanks{Corresponding author: \href{mailto:matteo@usm.lmu.de}{matteo@usm.lmu.de}},
E.~Rozo$^{2}$\thanks{Corresponding author: \href{mailto:erozo@email.arizona.edu}{erozo@email.arizona.edu}},
E.~S.~Rykoff$^{3,4}$,
A.~Farahi$^{5}$,
T.~Jeltema$^{6}$,
A.~E.~Evrard$^{6}$,
\newauthor
A.~Mantz$^{7,3}$,
D.~Gruen$^{3,4}\ddagger$,
R.~Mandelbaum$^{8}$,
J.~DeRose$^{7,3}$,
T.~McClintock$^{2}$,
T.~N.~Varga$^{1,9}$,
\newauthor
Y.~Zhang$^{11}$,
J.~Weller$^{1,9,10}$,
R.~H.~Wechsler$^{7,3,4}$,
M.~Aguena$^{12}$
\\
$^{1}$ Universit\"ats-Sternwarte, Fakult\"at f\"ur Physik, Ludwig-Maximilians Universit\"at M\"unchen, Scheinerstr. 1, 81679 M\"unchen, Germany\\
$^{2}$ Department of Physics, University of Arizona, Tucson, AZ 85721, USA\\
$^{3}$ Kavli Institute for Particle Astrophysics \& Cosmology, P. O. Box 2450, Stanford University, Stanford, CA 94305, USA\\
$^{4}$ SLAC National Accelerator Laboratory, Menlo Park, CA 94025, USA\\
$^{5}$ Department of Physics, University of Michigan, Ann Arbor, MI 48109, USA\\
$^{6}$ Santa Cruz Institute for Particle Physics, Santa Cruz, CA 95064, USA\\
$^{7}$ Department of Physics, Stanford University, 382 Via Pueblo Mall, Stanford, CA 94305, USA\\
$^{8}$ McWilliams Center for Cosmology, Department of Physics, Carnegie Mellon University, Pittsburgh, PA 15213, USA\\
$^{9}$ Max Planck Institute for Extraterrestrial Physics, Giessenbachstrasse, 85748 Garching, Germany\\
$^{10}$ Excellence Cluster Universe, Boltzmannstr. 2, D-85748 Garching, Germany\\
$^{11}$ Fermi National Accelerator Laboratory, P. O. Box 500, Batavia, IL 60510, USA\\
$^{12}$ Laboratorio Interinstitucional de e-Astronomia - LineA, Rua General Jose Cristino, 77, Rio de Janeiro, RJ, 20921-400, Brazil\\
$^{\ddagger}$ Einstein Fellow
}

\usepackage{eso-pic}

\AddToShipoutPictureBG*{%
  \AtPageUpperLeft{%
    \hspace{0.75\paperwidth}%
    \raisebox{-3.5\baselineskip}{%
      \makebox[0pt][l]{\textnormal{DES-2017-0318}}
}}}%

\AddToShipoutPictureBG*{%
  \AtPageUpperLeft{%
    \hspace{0.75\paperwidth}%
    \raisebox{-4.5\baselineskip}{%
      \makebox[0pt][l]{\textnormal{FERMILAB-PUB-18-325}}
}}}%

\begin{document}
\label{firstpage}
\pagerange{\pageref{firstpage}--\pageref{lastpage}}
\maketitle
\begin{abstract}

The cosmological utility of galaxy cluster catalogues is primarily limited by our ability to calibrate the relation between halo mass and observable mass proxies such as cluster richness, X-ray luminosity or the Sunyaev-Zeldovich signal. Projection effects are a particularly pernicious systematic effect that can impact observable mass proxies; structure along the line of sight can both bias and increase the scatter of the observable mass proxies used in cluster abundance studies.  In this work, we develop an empirical method to characterize the impact of projection effects on \redmapper\ cluster catalogues.  We use numerical simulations to validate our method and illustrate its robustness.  We demonstrate that modeling of projection effects is a necessary component for cluster abundance studies capable of reaching $\approx 5\%$ mass calibration uncertainties (e.g. the Dark Energy Survey Year 1 sample).  Specifically, ignoring the impact of projection effects in the observable--mass relation --- i.e. marginalizing over a log-normal model only --- biases the posterior of the cluster normalization condition $S_8 \equiv \sigma_8 (\Omega_{\rm m}/0.3)^{1/2}$ by $\Delta S_8 =0.05$, more than twice the uncertainty in the posterior for such an analysis.
\end{abstract}

\setcounter{footnote}{1}



\section{Introduction}
\label{sec:intro}

Galaxy clusters have played a significant role in the definition of the ``concordance'' \LCDM model \citep[for reviews, see e.g.][]{Allen2011,Kravtsov2012}. 
Current and upcoming wide-area photometric surveys --- e.g. the Dark Energy Survey (DES)\footnote{https://www.darkenergysurvey.org}, the Hyper
Suprime-Cam Subaru Strategic Program\footnote{http://hsc.mtk.nao.ac.jp/ssp/}, the Large Synoptic Survey
Telescope\footnote{https://www.lsst.org/}, Euclid\footnote{http://sci.esa.int/euclid/},
and {\it WFIRST}\footnote{https://wfirst.gsfc.nasa.gov/index.html} ---
seek to use the abundance and spatial distribution of galaxy clusters to improve constraints on the dark energy and the late-time normalization of the matter power spectrum.

One of the main limitations for the exploitation of galaxy clusters as cosmological tools is our ability to model the observable features of the massive halo population \citep[e.g.][]{Vikh2009,Rozo2010,Mantz2015,PlanckSZ2016}.  The observable mass proxy of interest within the context of the photometric surveys mentioned above is cluster richness. While the precise definition of cluster richness varies from catalogue to catalogue \citep[e.g.][and many others]{gladdersyee00,milleretal05,Hao2010,soares-santosetal11,Bellagamba2018}, in general cluster richness is a measure of galaxy content --- possibly weighted by luminosity --- of a galaxy cluster. In this work, we will focus specifically on cluster richness as defined in the red sequence Matched-filter Probabilistic Percolation algorithm \citep[redMaPPer;][]{Rykoff2014}.  This choice reflects both the excellent performance of \redmapper\ in the Sloan Digital Sky Survey data release 8 \citep[SDSS DR8,][]{Aihara2011,rozorykoff14}, and the fact that \redmapper\ is the cluster finding algorithm currently employed by the Dark Energy Survey (DES) collaboration \citep{Rykoff2016}.

As suggested by the name, \redmapper\ detects clusters as overdensities of red sequence galaxies.  
\redmapper\ estimates the probability that each red galaxy is a cluster
member using a matched filter approach, and then calculates the richness as the 
sum of the membership probabilities of all galaxies in the cluster field.  The sum extends
over all red-sequence galaxies above a fixed
luminosity threshold, and within an empirically calibrated cluster radius.
In order to maximize the cosmological utility of the \redmapper\ cluster sample,
the cluster richness defined by \redmapper\ has been optimized to 
minimize the scatter in the richness-mass relation \citep{Rozo2009,rozoetal11,Rykoff2012}.  \citet{rozoetal11} performed
an early study of systematic uncertainties affecting the richness estimates using the algorithm
employed by \redmapper.  Among the systematics studied in that work, two stood out: cluster miscentering,
and projection effects. Miscentering in \redmapper\ clusters has been studied in previous works \citep{sadibekovaetal14,rozorykoff14,hoshinoetal15,hikageetal17}, and additional work is on-going (Zhang et al., in preparation; von der Linden et al., in preparation).
Here, we focus exclusively on projection effects. 

Projection effects refer to the impact that correlated and uncorrelated structures along the line of sight can have on photometric cluster richness estimates (or any other observable mass proxy). In particular, the width of the red sequence, along with photometric uncertainties, places an inherent limit to the resolution that a photometric cluster finding algorithm can achieve along the line of sight \citep{Cohn2007}. Consequently, one expects richness estimates to be contaminated by the galaxy content of nearby structures.  Indeed, there are now multiple sources of observational evidence for projection effects in the SDSS \redmapper\ cluster catalogue \citep{farahietal16,zuetal17,buschwhite17}.  As emphasized in \citet{Erickson2011}, a detailed, quantitative characterization of these projection effects is necessary to successfully utilize galaxy clusters as a dark energy probe. This work seeks to establish the modeling framework necessary to quantify these effects for the SDSS and DES \redmapper\ cluster catalogues. 

A quantitative characterization of projection effects in a cluster catalogue faces two distinct challenges.  First, while one could imagine randomly inserting synthetic data clusters into the survey data set to study the impact of projection effects, any conclusions derived from such a study would not account for the impact of correlated large-scale structure around galaxy clusters. 
Conversely, any conclusions from simulation-based studies of projection effects in which galaxies are painted on dark matter halos will be limited by  uncertainties in the halo occupation distribution and galaxy color assignment used in the simulation \citep[see e.g.][]{vanHaarlem1997,Gerke2005,Cohn2007,farahietal16}.  Here we demonstrate how we can combine both real data and numerical simulations to tackle these twin challenges.  In particular, we rely on an analysis of real data to estimate the effect of background subtraction uncertainties and the magnitude of projection effects from uncorrelated large-scale structures.  At the same time, and exploiting the empirical understanding of projection effects gained from characterizing the impact of projection effects around random points in the SDSS data set,  we make use of mock catalogues to characterize the effects of correlated structures. While the method proposed here remains model-dependent --- as is necessarily the case for any simulation-based approach ---  our method has the virtue of being explicitly data-driven.  Moreover, the simplicity of our analysis enables multiple robustness tests that help us characterize the sensitivity of cosmological posteriors to our model assumptions.

The modeling framework detailed in this work will be utilized to derive cosmological constraints from the SDSS \redmapper\ catalogue (Costanzi et al. in preparation), and will be used by the DES collaboration in their upcoming analysis of the DES Year 1 \redmapper\ data set.  We also note that while the analysis in this paper is focused specifically on the SDSS \redmapper\ cluster catalogue, the algorithm developed here can be used to characterize projection effects in any cluster catalogue, including catalogues selected in other wavelengths.

The paper is organized as follows. In Section \ref{sec:method} we introduce the parametric model we have adopted for characterizing the impact of projection effects on the richness of galaxy clusters. Section \ref{sec:main} is devoted to the calibration and validation of our model.  Section \ref{sec:cosmo_anal} demonstrates that the work carried out in this paper is necessary for enabling accurate and precise cosmological inferences from the analyses of the abundance of \redmapper\ galaxy clusters.  We summarize and conclude in Section \ref{sec:conc}.


\section{Overview of the Projection Effects Model}
\label{sec:method}

Let $\lambda\ob$ denote the observed richness of a galaxy cluster.  The expectation value of the
density of galaxy clusters is given by 
\begin{equation}
\label{eqn:dndz}
 \langle n(\lambda\ob,z) \rangle = \int_{0}^{\infty} \de M\ n(M,z) 
 P(\lambda\ob | M,z) \, ,
\end{equation}
where $n(M,z)$ is the halo mass function, and $P(\lambda\ob | M,z)$ denotes the probability that a halo of mass $M$ at redshift $z$ is observed with richness $\lambda\ob$.  It is worth noting that this equation explicitly assumes that halos can be uniquely matched to clusters, which need not always be the case \citep[see e.g.][]{gerkeetal05}.  Using \redmapper\ clusters identified in simulated galaxy catalogues, \citet{farahietal16} demonstrate that 99\% of clusters with $\lambda\geq 20$ map to a unique dark matter halo while 1\% map to halos previously assigned to a richer system.  

The observed richness assigned to each cluster can be seen as the result of a two-step process:
first, the cluster has an inherent ``true'' richness, $\lambda\true$, which can be thought of as the richness the cluster finder would assign to an object in the absence of projection effects and observational errors. $\lambda\true$ is a random variable that depends on cluster mass.  Second, projection effects and photometric and observational noise perturb the richness $\lambda\true$ to arrive at the observed richness $\lambda\ob$ assigned to that galaxy cluster.  We can parametrize these stochastic contributions by decomposing $P(\lambda\ob | M,z)$ into a convolution of two distinct probability distributions:
\begin{equation}
\label{eqn:Plobmz}
 P(\lambda\ob | M,z) = \int_0^{\infty} \de \lambda\true\ P(\lambda\ob | \lambda\true,z) P(\lambda\true | M,z) \, ,
\end{equation}
where $P(\lambda\true | M,z)$ describes the intrinsic scatter of the richness--mass relation, and $P(\lambda\ob | \lambda\true)$ accounts for the additional scatter introduced by the observation\footnote{Here we are implicitly assuming that $\lambda\ob$ is independent of mass at fixed $\lambda\true$ and redshift. This assumption is validated a posteriori using our synthetic data, i.e.~by comparing $P(\lambda^{\rm ob}|M,\lambda\true,z)$ to $P(\lambda^{\rm ob}|\lambda\true,z)$ in our mock catalogs.}.  In this way, we disentangle any biases introduced by the cluster finder and the characteristics of the survey (e.g.~photo-$z$ uncertainty) from the underlying observable--mass relation.  The aim of this work is to provide a general procedure to calibrate $P(\lambda\ob | \lambda\true,z)$, and to demonstrate our method using the SDSS \redmapper\ catalogue. 

We model the perturbation of the observed richness relative to the true richness as the sum of two uncorrelated stochastic terms,
\begin{equation}
\label{eqn:lobNoP}
	\lambda\ob = \lambda\true + \Delta^{\rm bkg} + \Delta^{\rm prj} \, ,
\end{equation}
one due to photometric noise and the impact of observational uncertainties in the background subtraction,
$\Delta^{\rm bkg}$, and one which accounts for the effects of chance projections, $\Delta^{\rm prj}$.
The key distinction here is that $\Delta^{\rm bkg}$ is non-zero even when detecting clusters in unstructured background.  
By contrast, $\Delta^{\rm prj}$ refers to the contribution to $\lambda\ob$ from member galaxies of other halos projected along the line of sight.

The properties of the observational noise $\Delta^{\rm bkg}$ can be estimated directly from the data
by injecting synthetic galaxy clusters of known richness $\lambda\true$
into the survey data.  The injected clusters are added at the catalog level, not the image level.  When injecting clusters into the data, we fluctuate galaxy magnitudes according to the 
predicted magnitude errors given the local observing conditions, and then measure the 
observed richnesses $\lambda\ob$.  We detail our calibration
of this observational noise in Section~\ref{sec:method:rnd_data}.  One form of noise we do \it not \rm account for is noise due to centering failures. It is possible for projected clusters --- having an excess of galaxies relative to non-projected systems --- to be more likely to be miscentered than non-projected clusters.  That is, miscentering and projections effects might be correlated.  Because miscentering is well constrained (Zhang et al., in preparation) and has little impact on weak lensing mass \citep{desy1wl}, we expect any such corrections to be small, and postpone an investigation of this possibility to future work. As will be discussed in the cosmological analyses of the SDSS and DES \redmapper\ cluster samples, our approach is to correct the data for the effects of miscentering, rather than forward modeling the impact of miscentering on the data.\footnote{Note the weak lensing masses used in our analyses rely on forward modeling the impact of miscentering.  The recovered mass is used as the observable data vector for the cosmology analysis, and in that sense it is ``corrected for miscentering'', though the ``correction'' comes about from a forward-modeling treatment of the data.}  Consequently, the analysis in this work (which ignores miscentering) is directly applicable to those data sets.  In the future, we intend to forward model miscentering in addition to projection effects.

Unlike the background term $\Delta^{\rm bkg}$, we may not calibrate $\Delta^{\rm prj}$ through the injection of synthetic galaxy clusters into the data because galaxy clusters are not randomly distributed within the survey footprint; they live in over-dense regions, and correlated large-scale structure will boost projection effects relative to
estimates based on placing synthetic galaxy clusters at random points.  
To overcome this difficulty, we rely on N-body simulations, which allow us to place galaxy clusters at locations of massive dark matter halos.

At first glance, one might expect that to calibrate the impact of correlated structure one need only to populate $N$-body simulations with galaxies, and then run the \redmapper\ algorithm on the resulting mock galaxy catalogues.
The problem with
such an approach is that the projection effects depends in detail on how galaxies are distributed, 
particularly the color--redshift relation of red sequence galaxies: wider red sequences
will increase projection effects.  Consequently, an accurate calibration of projection effects requires 
a quantitatively accurate reproduction of not just the halo occupation distribution of cluster galaxies, but
also the red sequence width as a function of redshift for the mock galaxy catalogues.
 
To ensure that projection effects in our simulated data sets correctly mirror projection effects
in the real data we proceed as follows: given a halo catalogue in a light cone, we assign to each halo a richness $\lambda\true$.  The observed (i.e. projected) richness of a galaxy cluster is summing the 
true richnesses of all halos along the line of sight, weighted by a redshift kernel $w(\Delta z$), and the
fractional overlap area.
The kernel $w(\Delta z)$ characterizes the fractional contribution of the richness of a halo along the line
of sight to the projected richness of the dominant clusters.  Thus, $w=1$ when  
$\Delta z=0$ and $w=0$ when $|\Delta z|$ is large, i.e. halos separated by a large
redshift offset do not project onto each other.  
The richness perturbation due to projections takes the form
\begin{equation}
\label{eqn:prj}
\Delta^{\rm prj}_i= \sum_{j\neq i}^N  f^A_{ij} w(\Delta z_{ij}, z_j) \lambda\true_j \,,
\end{equation}
where the sum is over all clusters $j$ in the catalogue.
The coefficient $f^A_{ij}$ is a geometric term that accounts for misalignments between halos:
that is, projection effects should increase as the projected halo and the central
parent halo become more aligned.  At perfect alignment, $f^A_{ij}$ is equal to one. In the case of partial alignment, the fractional overlapping area is computed
analytically based on the radial offset between the parent and projected halos.
The key remaining task at this point is to specify the form of the filter function $w(\Delta z,z)$. 
The calibration of the filter function $w(\Delta z,z)$ is one of the key innovations in this work,
and is described in Section~\ref{sec:method:z_kernel}.

We seek to test the validity of our model for implementing projection effects on a simulation.  To this end, it is important to note that when we place synthetic galaxy clusters at random points in the survey, projection
effects still occur, even though they are suppressed due to the absence of correlated large-scale structure.
This enables a non-trivial test of the projection effects model: the projection effects of randomly placed
synthetic clusters in the simulation should match the projection effects observed in randomly placed
synthetic galaxy clusters in the real data.  That is, the probability distribution $P(\lambda\ob|\lambda\true,z)$
for randomly injected clusters in the simulation should match the corresponding distribution recovered from the data.
If the two distributions match, we can assert that our method for implementing projection effects in the simulation correctly mirrors how projection effects occur in the real data.  This validation test is performed in Section~\ref{sec:method:mock_val}.

Having validated our methodology for incorporating projection effects into $N$-body simulations, 
in Section~\ref{sec:prj_LSS} we characterize the impact of projection effects on richness estimates including correlated large-scale structure, i.e. we re-calibrate the probability distribution of the richness perturbation $\Delta^{\rm prj}$ considering clusters at their actual position within the large-scale structure of the simulated universe.

In short, our analysis ultimately relies on two types of synthetic data sets: 1) synthetic galaxy clusters that are injected into the real SDSS data in order to quantify projection effects along random points.  2) N-body simulations in which halos are assigned true richnesses, which are then analytically projected to arrive at an observed (projected) richness for each halo.

It is worth reiterating that the work described here does {\it not} rely on populating $N$-body simulations with galaxies, and then running \redmapper\ on the resulting cluster catalogue.  Doing so would enable us to calibrate projection effects in the simulations, but the results are necessarily dependent on the details of the input color distribution and evolution of the galaxies.  By contrast, our empirically motivated analysis enables us to calibrate projection effects in a controlled, data-driven way.  In future work, we intend to apply the methods described here to \redmapper\ runs on synthetic galaxy populations to further validate our methodology on synthetic universes.


\section{Analysis}
\label{sec:main}

Here we:
\begin{itemize}
\item Calibrate the noise in the cluster richness estimates associated with photometric noise
and stochasticity in the background galaxy population (Section~\ref{sec:method:rnd_data}).  We also characterize projection effects due to uncorrelated large scale structure using synthetic clusters at random points in the real data.  
\item Describe how we introduce projection effects into simulated halo catalogs to arrive at synthetic cluster catalogs that include projections (Section~\ref{sec:method:z_kernel}).
\item Validate our model by comparing the incidence of projection effects for randomly located 
synthetic clusters in the simulation to that of the data (Section~\ref{sec:method:mock_val}).  
\item Calibrate the incidence of projection effects in the simulation including correlated
large-scale structure (Section~\ref{sec:prj_LSS}).
\end{itemize}

Throughout the paper quantities labeled with ``RND'' are derived using randomly located synthetic clusters, while quantities labeled with ``LSS'' are derived using synthetic clusters placed at the appropriate halo positions within the large-scale structure of the simulation.  That is, ``LSS'' quantities properly account for the impact of correlated structures.  

\subsection{Calibration of Observational Noise and Projection Effects from Uncorrelated
Large Scale Structure}
\label{sec:method:rnd_data}

To characterize observational noise we inject synthetic clusters at random positions in the sky,
and compare the recovered richness $\lambda\ob_{\rm out}$ to the true input richness of the synthetic
clusters $\lambda\true_{\rm in}$.
Our synthetic galaxy clusters are generated using an improved version of the method outlined in \cite{Rykoff2014}, which makes use of the red sequence color model calibrated from the data and depth maps. 
In brief, given a true cluster richness and redshift $(\lambda\true_{\rm in},z_{\rm in})$ we proceed as follows:
\begin{enumerate}
	\item{First, we generate a list of $10,000$ random positions, uniformly sampling the survey mask.}
    \item{At each location, we place $\lambda\true_{\rm in}$ galaxies distributed in radius and color--magnitude space according to the empirically calibrated red-sequence model of \redmapper.  The magnitudes of the cluster galaxies are then perturbed according to the expected photometric noise as reported in the SDSS depth maps.  The red-sequence calibration assumes a linear model in color-magnitude space, with a multivariate Gaussian scatter for the photometric magnitudes. The expectation value and covariance matrix characterizing the red-sequence model are iteratively trained on spectroscopic clusters using a maximum-likelihood method for estimating the parameters.  For further details, we refer the reader to \citep{Rykoff2014}. While our methodology does not insert blue galaxies -- i.e. galaxies not described by the red-sequence model detailed above -- whether in the cluster or otherwise, these galaxies have zero weight when computing cluster richness, and therefore do not impact the performance of the cluster finder.
}
    \item{We measure the richness $\lambda\ob$ of the synthetic galaxy clusters.}
\end{enumerate}
The procedure above does not account for miscentering errors, nor for ``catalogue noise'', i.e. the stochasticity associated with galaxy detections.  In our upcoming cosmological analyses, the impact of cluster miscentering on the cluster number counts and weak lensing mass estimates is explicitly accounted for by correcting the observed data vectors (abundance and weak lensing masses) for the effects of miscentering.  As for stochasticity due to catalogue noise, any such stochasticity will necessarily be subsumed into estimates of the intrinsic scatter of the richness--mass relation when performing cosmological analyses with the \redmapper\ cluster samples.

As detailed in \cite{Rykoff2014} \redmapper\ analyzes clusters in three stages. First it looks for overdensities of red-sequence galaxies. Second, for every cluster of galaxies, it computes the probability for each galaxy to be a cluster member.  Third, after sorting the cluster candidates according to the cluster likelihood, it percolates through the full catalogue while probabilistically masking out cluster members.  In the interest of simplicity, in this first pass we will ignore the impact of percolation, which only affects a small percentage of the clusters.  However, we return to characterize the impact of percolation on $P(\lambda\ob | \lambda\true,z)$ in Section \ref{sec:prj_LSS}.
With this simplification, the galaxies of our synthetic clusters are never absorbed by higher richness systems, and therefore projection effects can only increase the observed richness of synthetic clusters.

\begin{figure*}
\begin{center}
    \includegraphics[width=0.65 \textwidth]{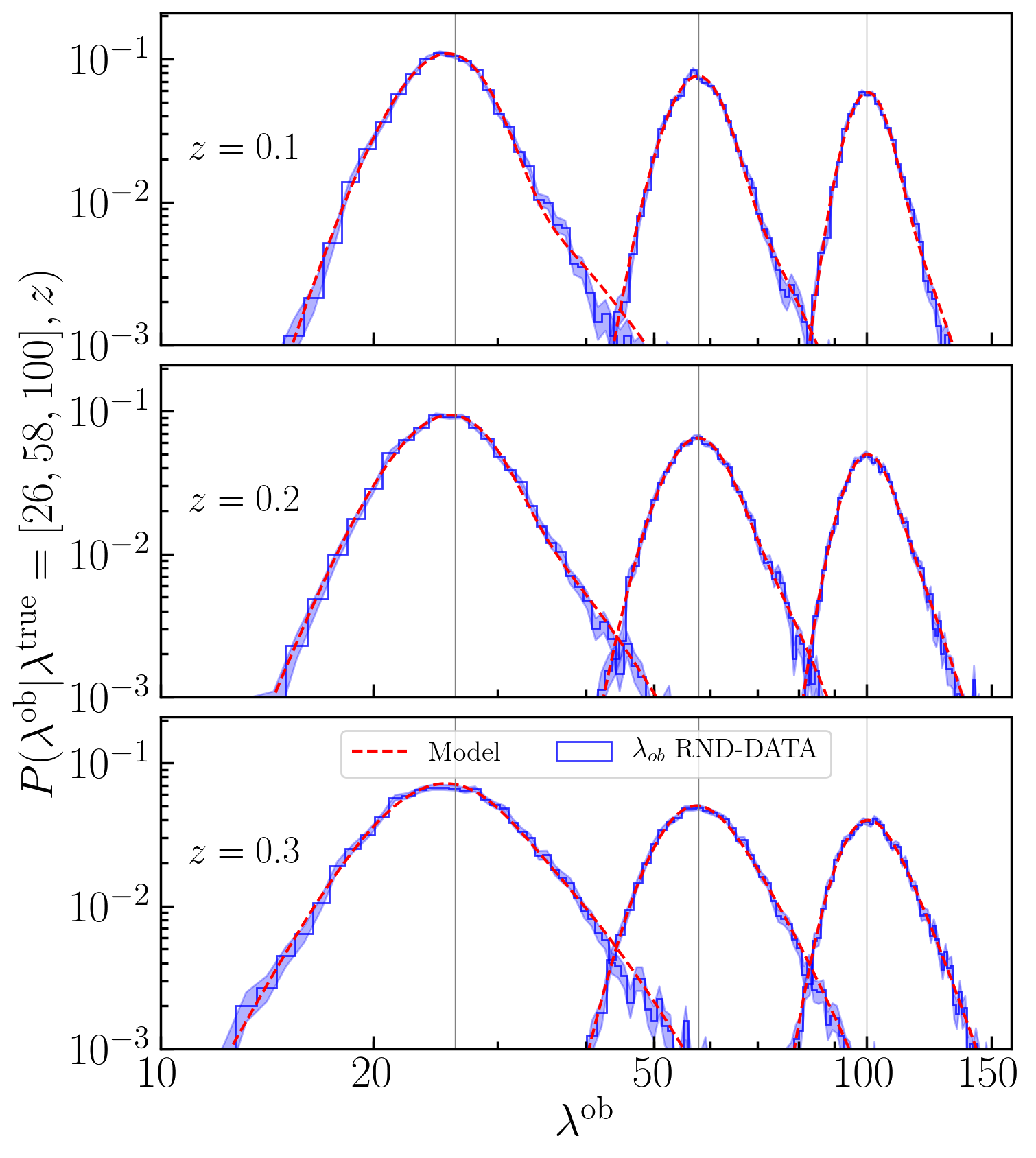}
\end{center}
\caption{The blue histograms show the 
probability distribution $P(\lambda\ob | \lambda\true,z)$ obtained by randomly injecting
synthetic galaxy clusters into the SDSS data for a grid of input redshifts and richnesses as
labeled. They show the effect of photometric noise and projection effects, yet without the additional scatter due to correlated structure and percolation. The solid-red lines are the best fit analytic models (\autoref{eqn:Prnd}) to these distributions. The vertical lines correspond to the $\lambda\true$ values used to generate the distributions.
}
\label{fig:rnds}
\end{figure*}

For our analysis we inject synthetic clusters into the SDSS DR8 data.  The richness and redshifts of the
injected clusters are taken from a grid along these two axis, with 
$\lambda\true_{\rm in}=[5,15,26,36,47,58,68,78,89,100]$ and $z_{\rm in}=[0.1,0.15,0.2,0.25,0.3]$.
The blue histograms in Figure \ref{fig:rnds} show the probability distributions $P(\lambda\ob | \lambda\true,z)$ recovered from our analysis for three different richness  and redshift bins as labeled.  As expected, the distributions are wider and more positively skewed for larger objects -- i.e. larger $\lambda\true$ -- and at higher redshift; larger objects have a larger cross section, increasing the chance of spurious projections, while higher redshift systems suffer from larger photometric errors.

\begin{figure}
\begin{center}
    \includegraphics[width=0.45 \textwidth]{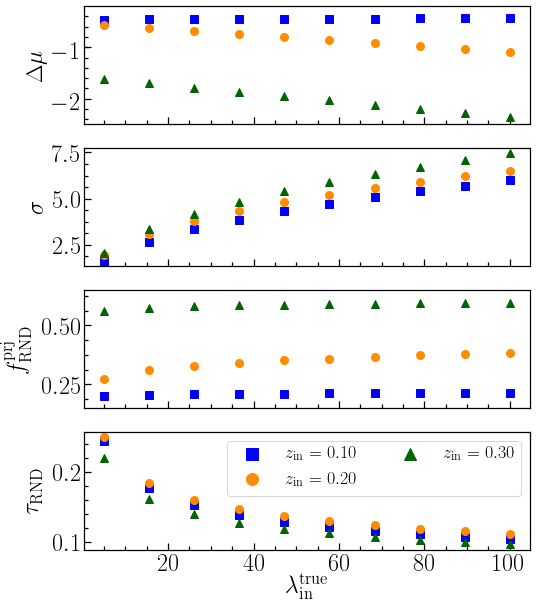}
\end{center}
\caption{Dependence on the input richness and redshift of the model parameters which characterize the observational scatter and the magnitude of projection effects due to uncorrelated structures  (Eq. \ref{eqn:Prnd}) .}
\label{fig:prj_par}
\end{figure}

We model the fluctuations in the observed richness as the sum of two stochastic perturbations, $\Delta^{\rm bkg}$ and $\Delta^{\rm prj}$, as per equation \ref{eqn:lobNoP}.
$\Delta^{\rm bkg}$ is assumed to be Gaussian distributed, $\mathcal{N}(\Delta\mu,\sigma)$, where both the bias in the recovered richness $\Delta\mu$ and the scatter $\sigma$ are functions of $(\lambda\true,z)$.  The distribution
$P(\Delta^{\rm prj})$ is modeled as the sum of an exponential and a delta distribution:
\begin{equation}
\label{eqn:P_dprj}
P(\Delta^{\rm prj} | \lambda\true,z) = (1-f^{\rm prj})\delta_D(\Delta^{\rm prj})+ f^{\rm prj} \tau e^{-\tau \Delta^{\rm prj}} \Theta(\Delta^{\rm prj}) .
\end{equation}
We have found empirically that this parametric model provides an accurate description of our simulated data. 
In the above expression, $f^{\rm prj}(\lambda\true,z)$ is the fraction of objects affected by projections, and the step function $\Theta(\Delta^{\rm prj})$ ensures $\Delta^{\rm prj}\geq 0$.  The parameter $\tau(\lambda\true,z)$, defining the steepness of the exponential distribution, characterizes the magnitude of projection effects. Note that $\tau$ has a unit of inverse richness, and small values of $\tau$ correspond to stronger projection effects.
The convolution of the two distributions is:
\begin{multline}
\label{eqn:Prnd}
P(\lambda\ob | \lambda\true,z)= (1-f^{\rm prj}) \mathcal{N}~(\mu,\sigma) + \\
+ f^{\rm prj} \frac{\tau}{2} \exp\left[ \frac{\tau}{2} (2\mu + \tau\sigma^2 - 2 \lambda\ob)\right] {\rm erfc} \left( \frac{\mu +\tau\sigma^2-\lambda\ob}{\sqrt{2}\sigma}\right) \, ,
\end{multline}
where we have defined $\mu=\lambda\true + \Delta\mu$.  This expression contains 4 independent parameters, $\{\Delta\mu,\sigma,f^{\rm prj},\tau\}$, which are fit by matching
our parametric model to the probability distributions $P(\lambda\ob | \lambda\true,z)$ recovered
from injecting clusters into the SDSS data sets. 
As shown by the {\it dashed red} lines in Figure \ref{fig:rnds}, our model provides a good fit to the
data. The dependence of the best fit parameters  on the
input richness and redshift is shown in Figure \ref{fig:prj_par}.  These parameters characterize the
impact of observational errors and the impact of 
uncorrelated large-scale structure in the observed cluster richness.
For uncorrelated structure, at fixed redshift, the magnitude of projection effects increases --- i.e. $\tau$ decreases --- as a function of $\lambda\true$ due to the larger angular area subtended by the richer cluster. 
The pronounced increase in the fraction of clusters with projection effects at $z\approx 0.3$ is expected: at lower redshifts, projections are primarily due to the non-zero width of the red-sequence.  Once photometric noise becomes larger than the intrinsic width of the red-sequence, the impact of projection effects increases with increasing redshift, leading to the enhancement seen in Figure~\ref{fig:prj_par}.

An important result from this analysis is that the richness errors 
quoted in the \redmapper\ catalogue underestimate the true observational uncertainty by $\sim 40\% - 70\%$ depending on the richness and redshift of the cluster.  This difference
is due to the fact that the errors quoted in the  \redmapper\ catalogue represent the statistical uncertainty in the
total number of cluster galaxies \it assuming the membership probabilities are correct. \rm  That is, it only accounts
for the stochasticity in cluster membership.  In practice, the membership probabilities themselves are subject to observational noise, boosting the observed error relative to the error calculated by \redmapper.

\subsection{How to Include Projection Effects in Simulated Data}
\label{sec:method:z_kernel}

We wish to develop a method for adding projection effects to simulated data
in a way that faithfully reproduces the impact of projection effects in the real data.
Populating halos with galaxies in a multi-dimensional color space in a way that adequately matches the detailed trends of cluster galaxies is difficult. Here, we assign richness values to halos according to a richness--mass relation, and then project and percolate the halo catalogue in a way that mirrors the \redmapper\ algorithm. 
We validate our method in section \ref{sec:method:mock_val}. 

Here, we describe our algorithm for computing the projected richness of a galaxy cluster given a simulated
data set in which all halos along the line of sight have been assigned an intrinsic richness.
Specifically, given a halo with known position, redshift, projected area and an assigned richness $\lambda\true$,
we assign a total \it projected \rm richness as follows: 
\begin{enumerate}
\item{Sort the mock halo catalogue according to $\lambda\true$ --- the assigned intrinsic richness --- in descending order. This step mimics the rank-ordering procedure of \redmapper.}
\item{Starting from the richest cluster in the catalogue, assign an observed (projected) richness according to 
\begin{equation}
\label{eq:prj}
\lambda\ob_i = \lambda\true_i + \Delta^{\rm prj}_i= \lambda\true_i + \sum_{j\neq i}^N \lambda\true_j f^A_{ij} w(\Delta z_{ij},z_j)  .
\end{equation}
Here, $\lambda\true$ is the richness of a galaxy cluster assigned using a fiducial scaling relation.
$f^A_{ij}$ is the fraction of area  of the $j$-th object which overlaps with the area  of the $i$-th object in projection.  This assumes that the galaxies are uniformly distributed inside the cluster radius; while a crude approximation, we have explicitly verified that using a more realistic radial profile model does not significantly impact the resulting analysis. We have not explored the sensitivity of our analysis to allowing for elliptical galaxy distributions, which we leave for future work.  Finally, $w(\Delta z_{ij}|z_j)$ is a redshift-dependent weight which accounts for the redshift distance between $i$ and $j$.  We detail below how the function $w(\Delta z)$ is calibrated.}
\item{For tests that include percolation effects, having measured the observed (projected) richness of cluster $i$, 
we update the intrinsic richness (and therefore the radius) 
of all clusters $j>i$ via $\lambda\true_j=\lambda\true_j(1- f^A_{ij} w(\Delta z_{ij},z_j))$.  This update subtracts out from each cluster $j$ the galaxies that this cluster contributed to a richer system, 
mirroring the percolation algorithm employed in \redmapper.  The richness perturbation $\Delta^{\rm prc}$ due to percolation can be written as:
\begin{equation}
\Delta^{\rm prc}_j= \sum_{i < j}^N \lambda\true_j (1- f^A_{ij} w(\Delta z_{ij},z_j))
\end{equation}
This quantity is a third source of stochastic noise that impacts the observed richness of a galaxy cluster, and is added to the $\Delta^{\rm prj}$ and $\Delta^{\rm bkg}$ contributions in equation~\ref{eqn:lobNoP}.
}
\item{We move from cluster $i$ to cluster $i+1$, and iterate until we move through the whole halo list, arriving at our synthetic cluster catalogue.}
\end{enumerate}

At this point, the richness $\lambda\ob$ does not include the noise due to observational errors, $\Delta^{\rm bkg}$. The final value for $\lambda\ob$ is obtained by adding a Gaussian random draw from the $\Delta^{\rm bkg}$ distribution calibrated in the previous section. 
When comparing our simulated data to the synthetic random clusters of section~\ref{sec:method:rnd_data} we do not apply the percolation step, but when calibrating the full impact of projection effects (Section~\ref{sec:prj_LSS}) we explicitly incorporate this effect.

\begin{figure*}
\begin{center}
    \includegraphics[width=\textwidth]{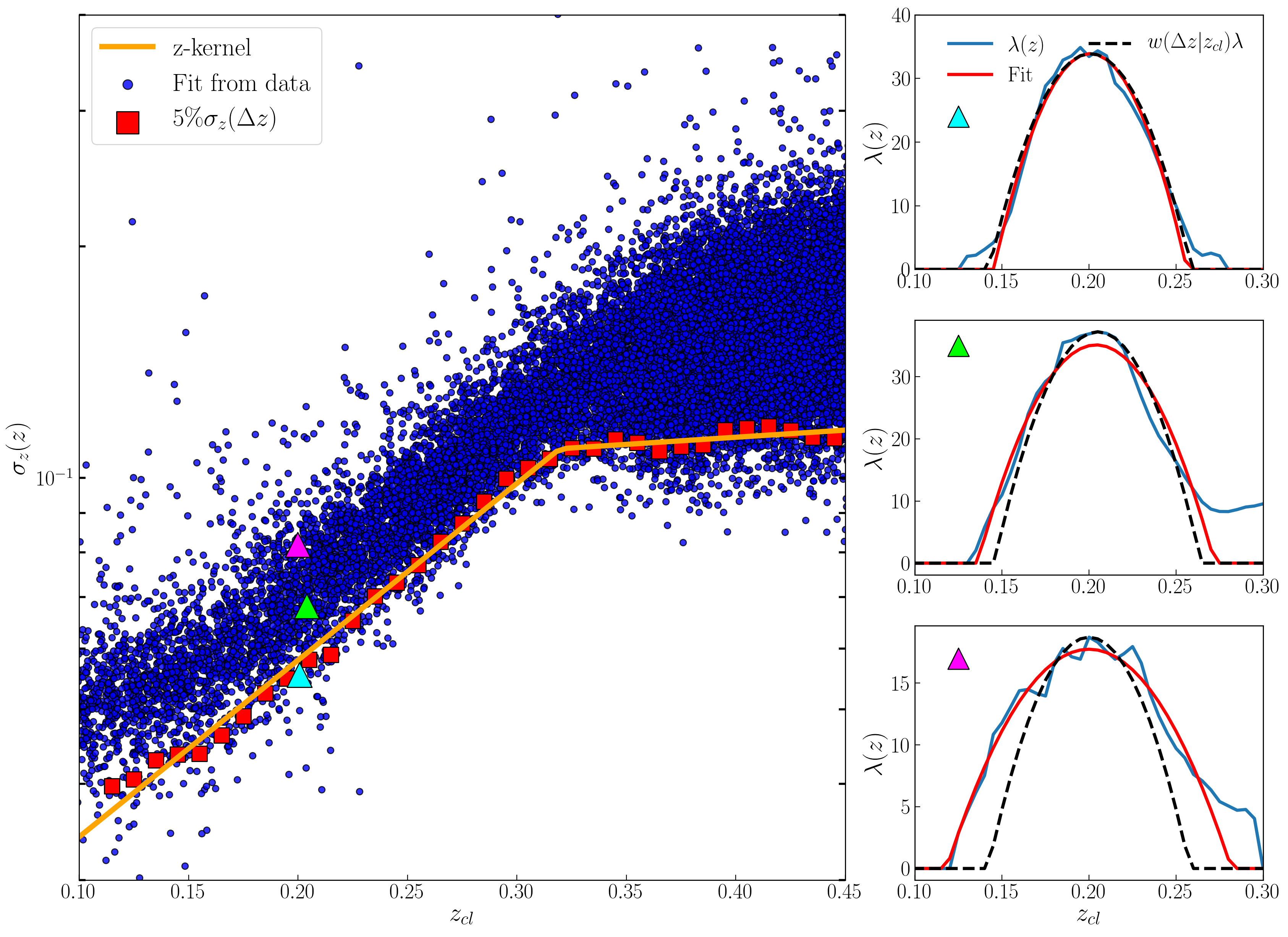}
\end{center}
\caption{{\it Left} panel: The blue dots are the best-fit values for $\sigma_z$ obtained when fitting the curves $\lambda(z)$ for each cluster in the \redmapper\ cluster catalogue (see text).
The {\it red squares} represent the $5$ percentile of the $\sigma_z$ distribution estimated in redshift bins of width $\Delta z=0.01$. The {\it solid orange} line shows the model for $\sigma_z(z)$ adopted for the analysis. {\it Right} panels: The {\it blue solid} lines are the measured $\lambda(z)$ for the three clusters labeled with triangles in the {\it left} panel; the {\it red solid} lines represent the best-fit model for $w(\Delta z,z_{\rm cl})$; for comparison, the {\it black dashed} lines show the redshift kernel expected for {\it clear} l.o.s. clusters, i.e. assuming the calibrated $\sigma_{z}(z)$ in the computation of the kernel $w(\Delta z,z_{\rm cl})$.}
\label{fig:z_kernel}
\end{figure*}

In order to fully specify our projection algorithm we must calibrate the function $w(\Delta z,z_{\rm cl})$, 
that is, the fraction of galaxies that a high richness cluster will absorb from a lower-ranked cluster at 
redshift $z_{\rm cl}+\Delta z$, assuming perfect alignment of the two systems. 
This function is specific to the survey and cluster finding algorithm under consideration, and must be calibrated directly from the data. For this calibration, we re-measure the richness of every cluster in the \redmapper\ catalogue along a grid of redshift values around each cluster's true redshift (see e.g. {\it right} panels of Figure \ref{fig:z_kernel}).  Given a cluster at redshift $z_{\rm cl}$, the richness $\lambda(z)$ gives us the number of galaxies that would ``leak'' into a higher-ranked object along the same line of sight at redshift $z$.  That is,
we expect the function $\lambda(z) = \lambda w(\Delta z,z_{\rm cl})$, where $\lambda$ is the true richness of the cluster, and $\Delta z = z-z_{\rm cl}$ is the redshift offset. Our analytic model for $w(\Delta z,z_{\rm cl})$ is:
 \begin{equation}
 \label{eqn:w_z}
 w(\Delta z,z_{\rm cl})= \left \{ \begin{array}{rl}1-\frac{(\Delta z)^2}{\sigma_{z}(z_{\rm cl})^2} \; , & |\Delta z|<\sigma_z(z_{\rm cl}) \\ 0 \; , & \mathrm{otherwise} \end{array} \right .
 \end{equation}
This functional form arises from the simple expectation that $w(\Delta z,z_{\rm cl})=1$ when $\Delta z = 0$, and that $w(\Delta z,z_{\rm cl})=0$ when $|\Delta z|$ is larger than some maximum separation $\sigma_z$, where $\sigma_z$ depends on the cluster redshift.  

Example fits to the function $w(\Delta z,z_{\rm cl})$ as measured in the SDSS data are shown in the right panels of Figure \ref{fig:z_kernel}.  Note that our functional form has only one free parameter, $\sigma_z$.  The best fit values for $\sigma_z$ for every object in the SDSS \redmapper\ sample are shown in the left panel of Figure \ref{fig:z_kernel}. As expected,  the size of the kernel increases with redshift due to larger photometric errors.  At $z>0.33$ the \redmapper\ catalogue is no longer volume limited and the faintest galaxies detected for every cluster reside at the survey's limiting magnitude.  This results in the roughly constant width of the kernel function $w(\Delta z,z_{\rm cl})$ at high redshifts. The scatter in $\sigma_{z}$ values between clusters at the same redshift reflects the presence or absence of structures along the line of sight to each cluster: secondary structures add their richness to the naive expectation $\lambda w(\Delta z,z_{\rm cl})$, thereby broadening the measured $\lambda(z)$ function.  Consequently, if we wish to measure the ``leakage'' $w(\Delta z,z_{\rm cl})$, we should restrict ourselves to clusters which reside along clean lines of sight, i.e. clusters for which there is no broadening of the curve $\lambda(z)$ due to structures along the line of sight. We estimate the leakage function $w(\Delta z,z_{\rm cl})$ as the lower envelope defining the 5\% narrowest kernels in Figure \ref{fig:z_kernel}.  This 5\% envelope is estimated in redshift bins of width $\Delta z=0.01$, which are fit with a broken log-linear model.  Our best fit, shown with the {\it orange} line in Figure \ref{fig:z_kernel}, gives the relation:
\begin{eqnarray}
\label{eqn:z_kernel}
\log \sigma_z(z) &=& 2.299\,( z -0.32) -0.961 \quad {\rm for} \, z\leq 0.32 \\ 
\log \sigma_z(z) &=& 0.185\, (z -0.32 ) -0.961 \quad {\rm for} \, z>0.32 \nonumber \, .
\end{eqnarray}
We have explicitly verified that our final model for projection effects is robust to modest modifications
of our method for calculating the lower envelope of the data for $\sigma_z(z)$ shown in Figure~\ref{fig:z_kernel}.
Specifically, we verified that using the $10^{\rm th}$ percentile of the $\sigma_z$ distribution to define $\sigma_z(z)$ does not appreciably affect our results (see section \ref{sec:method:mock_val} and \ref{sec:cosmo_anal}).
The break in the figure reflects the transition of the SDSS \redmapper\ catalog
from being volume limited to limited by the survey depth.

%


\subsection{Validation of the Projection Effects Model on Simulated Data}
\label{sec:method:mock_val}

We seek to validate our model for introducing projection effects in simulations by generating a
synthetic cluster catalogue, and testing whether the projection effects from uncorrelated large scale structure in this
mock catalog match the observational results from Section~\ref{sec:method:rnd_data}.  Agreement on the impact
of projection effects between the simulated and real data sets constitutes strong evidence that our methodology for including projection effects in the simulation is valid.

To generate a synthetic cluster catalogue, we start with the halo catalogue extracted from an $N$-body 
simulation of a flat-\LCDM cosmological model with 
$\Omega_{\rm m} = 0.286$, $h_0 = 0.7$, $\Omega_{\rm b} = 0.047$, $n_s = 0.96$, and $\sigma_8 = 0.82$ 
(DeRose et al. 2018, in prep, Wechsler et al. 2018, in prep).
The simulation, containing $1400^3$ particles in a $[1050\ h^{-1}\ {\rm Mpc}]^3$ volume, has been run with the L-Gadget code, a variant of Gadget \citep{Springel2005}. 
A lightcone covering a quarter of the sky, over the redshift range $0.1 < z < 0.9$, was output from the simulation on the fly.
The halo catalogue has been created with the {\tt Rockstar} halo finder \citep{Behroozi2013} and it includes halos down to $M_{200m}=10^{12.5} [M_\odot / h]$.  Throughout, all masses refer to an overdensity of 200 with respect to the mean.
To assign a richness to the halos we rely on the results of \cite{Simet2016}, who placed constraints
on the mass--richness relation of SDSS \redmapper\ clusters.
Specifically, we assign to each halo a richness drawn from a log-normal distribution having mean:
\begin{equation}
\label{eqn:OMR}
  \ln \langle \lambda\true|M \rangle = \ln \lambda_0 + \alpha \ln \left(\frac{M}{M^*} \right) \, ,
\end{equation}
and variance:
\begin{equation}
\label{eqn:sig_lm}
 \sigma_{\ln \lambda-M}^2 = \frac{\langle  \lambda\true|M \rangle - 1}{\langle \lambda\true|M \rangle^2} + \sigma_{\rm intr}^2 \, ,
\end{equation}
where the model parameter values are: $\alpha=0.70$, $\lambda_0=40$, $\log M^*=14.348$ and $\sigma_{\rm intr}=0.25$. 
The scatter model is Poissonian when the number of satellite galaxies is low, but super-Poissonian at high occupancy.
The fiducial value of the scatter parameter $\sigma_{\rm intr}$ is motivated from comparisons of the \redmapper\ catalogue to X-ray and SZ clusters \citep{rozorykoff14,rozoetal15}. A more extensive analysis of the scatter of the richness--mass relation from comparison to X-ray data will be presented in an upcoming paper (Farahi et al., in preparation). Finally, as per the convention adopted by the \redmapper\ algorithm, we assign a physical radius to each halo based on its assigned richness: $R(\lambda)=\left( \lambda/100 \right)^{0.2}  [{\rm Mpc}/h]$. The radius, in turn, defines the projected angular extent of the halo: $\pi [R_\lambda /D_A(z)]^2$.

Given this simulated cluster catalogue, we assign a projected cluster richness to every halo as detailed in section~\ref{sec:method:z_kernel}.  To validate the projected richnesses, we inject 5000 clusters in the simulated data set at random positions, and compute their projected cluster richness.  These injected clusters are simply tagged with a richness value, not a full galaxy distribution, and therefore there is no ``observational noise'' associated with the injection.  Instead, we add the Gaussian random noise calibrated in Section~\ref{sec:method:rnd_data} to the observed richness for each cluster.  Since this noise is Gaussian, any non-Gaussian tails in the simulated data set necessarily come from the projection effects modeling described in section~\ref{sec:method:z_kernel}.

\begin{figure}
\begin{center}
    \includegraphics[width=0.50 \textwidth]{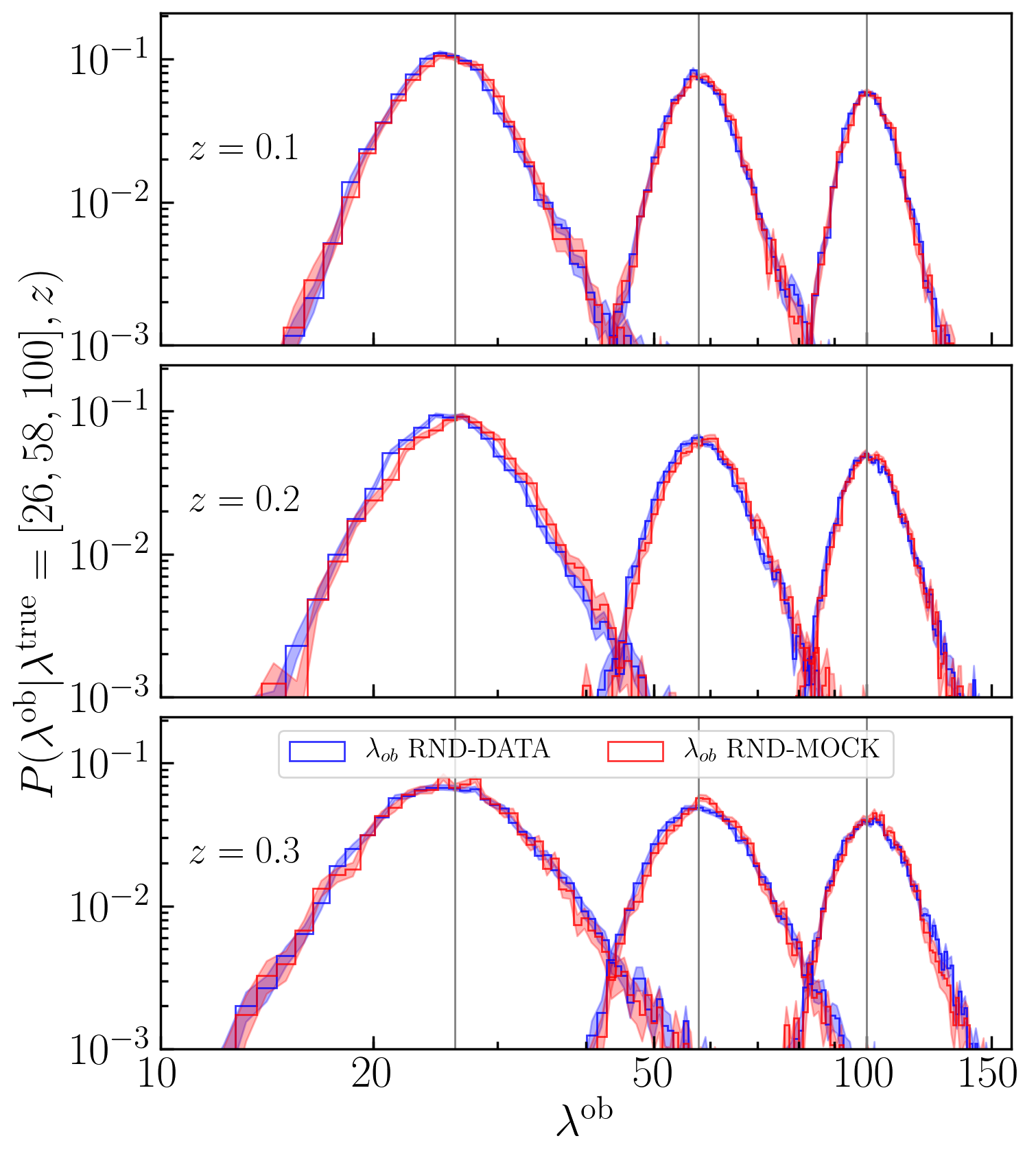}
\end{center}
\caption{Comparison of the probability distributions $P(\lambda\ob | \lambda\true,z)$ as recovered from the data ({\it blue} histograms) and our simulated catalogue ({\it red} histograms).  In both cases, synthetic clusters are added at random positions when measuring $\lambda\ob$.  The shaded areas correspond to the statistical uncertainty of the samples.
}
\label{fig:data_vs_mock}
\end{figure}

Figure~\ref{fig:data_vs_mock} compares the probability distributions $P(\lambda\ob|\lambda\true,z)$ recovered in our simulated data set (red histogram) to the SDSS results obtained by injecting synthetic clusters at random positions (blue histogram). The shaded regions for the simulated data set correspond to the uncertainty in the measurement from our simulations. There is excellent agreement between the two distributions at all input richnesses and redshifts tested. A small horizontal shift of the distributions can be seen in the middle panel of Figure~\ref{fig:data_vs_mock}.  However, the {\it shape} of the distributions is exceedingly well matched, and small horizontal shifts are exactly degenerate with the richness--mass relation, and therefore trivially absorbed into the nuisance parameters of a standard cluster abundance study.

\begin{figure}
\begin{center}
    \includegraphics[width=0.50 \textwidth]{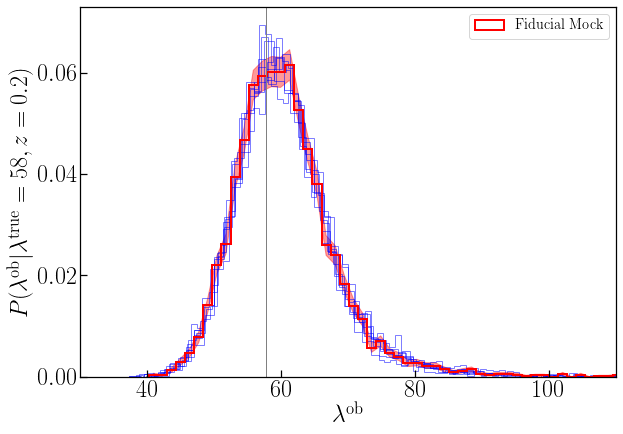}
\end{center}
\caption{The probability distributions $P(\lambda\ob | \lambda\true,z)$ as calibrated under a variety of different assumptions. The {\it red} line shows the distribution recovered from our fiducial mock catalogue detailed in section \ref{sec:method:mock_val}. The {\it blue} histograms are obtained from mock catalogues in which we vary
the richness-mass relation parameters within their allotted errors, or by using the $10$ percentile of the $\sigma_z$ distribution to calibrate the redshift kernel. 
These differences have only a modest impact on the resulting probability distribution 
$P(\lambda\ob|\lambda\true,z)$.}
\label{fig:PDFvsSyst}
\end{figure}

A key question is the degree to which the recovered incidence of projection effects depends on the details of the input richness--mass relation used in our analysis.   
We have verified that shifting the richness--mass relation parameters used to generate the mock catalogue within the $1\sigma$ error quoted in \cite{Simet2016}, including the intrinsic scatter in richness, has a negligible effect on the resulting distributions. 
Likewise, modest changes in the calibration of the redshift kernel --- e.g. considering the $10^{\rm th}$ percentile of the $\sigma_z$ distribution instead of the $5^{\rm th}$ percentile to define $\sigma_z(z)$ --- does not appreciably affect our results. As an example, an explicit comparison of these variations is shown in Figure~\ref{fig:PDFvsSyst} for clusters of input richness $\lambda\true=58$ at $z=0.2$.

\begin{figure}
\begin{center}
    \includegraphics[width=0.50 \textwidth]{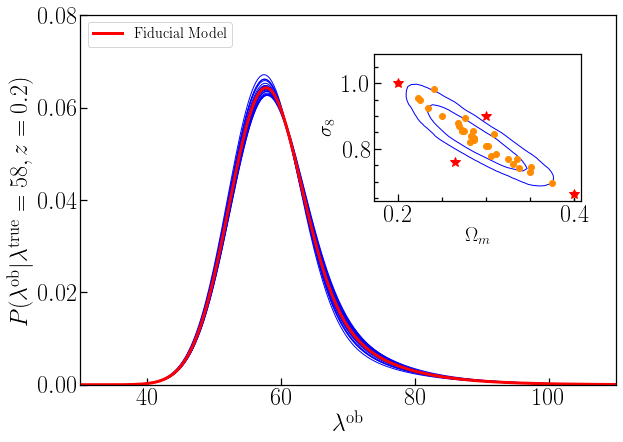}
\end{center}
\caption{Effect of cosmology on the probability distributions $P(\lambda\ob | \lambda\true,z)$. The {\it red solid} line show our best-fit model to the data. The {\it blue} lines are obtained shifting the best-fit values of $\tau$ and $f^{\rm prj}$ according to the analytically derived correction detailed in Appendix \ref{app:prj_eff} for different cosmologies. The orange dots in the inset plot show the cosmological parameter values used to generate the thin blue lines in the above Figure.  The confidence contours of the inset are those derived from the mock cosmological analysis of Section \ref{sec:cosmo_anal}. The red stars correspond to the cosmologies tested in Section \ref{sec:cosmo_anal} to assess the sensitivity of our cosmological posteriors to the input cosmological model used when calibrating $P(\lambda\ob | \lambda\true,z)$.}
\label{fig:PDFvsCosmo}
\end{figure}

Finally, the magnitude of projection effects should also depend on cosmology; e.g., larger values of $\sigma_8$ and/or $\Omega_m$ entail a larger number density of halos and thus stronger projection effects. In Appendix \ref{app:prj_eff} we develop an analytic model for calculating the relative shift of the parameters characterizing the projection effects -- $f_{\rm prj}$ and $\tau$ -- as a function of cosmology. 
Figure \ref{fig:PDFvsCosmo} shows the response of $P(\lambda\ob| \lambda\true)$ to the cosmological dependent shifts of the projection effects parameters. Specifically, the {\it blue} lines have been obtained from Equation \ref{eqn:Prnd} using the best-fit parameters derived from the data, but correcting $\tau$ and $f^{\rm prj}$ for the analytically derived shift corresponding to $20$ different cosmologies. The $20$ input cosmologies, shown with {\it orange} {\it dots} in the inset plot of Figure  \ref{fig:PDFvsCosmo}, have been chosen sampling the posterior distribution derived from the cosmology analysis of the simulated data set using our fiducial values (see section~\ref{sec:cosmo_anal}). This test shows that the cosmological sensitivity of the projection effects is mild.

We further demonstrate the robustness of our calibration to the input cosmology by selecting a few cosmological models that fall just outside the 95\% confidence region of the posterior derived from our simulated data set ({\it red stars} in the inset plot).  
We calculate $f_{\rm proj}$ and $\tau$ for these models, and explicitly verify that the resulting cosmological posteriors are not significantly different when using these values.  That is, the cosmological dependence of projection effects has a negligible impact on cosmological constraints obtained using our model.  For further details, see section~\ref{sec:cosmo_anal}.

\subsection{Characterization of the Projection Effects Due to Correlated large-scale Structure}
\label{sec:prj_LSS}

Having validated our algorithm to include projections and observational noise into our simulated cluster catalogues, we proceed to calibrate the full distribution $P(\lambda\ob | \lambda\true,z)$ for real galaxy clusters, including the impact of correlated large scale structure. We again begin by assigning to every halo in our simulation a true intrinsic richness $\lambda\true$ from the distribution $P(\lambda\true|M,z)$.  We wish to calibrate the distribution $P(\lambda\ob|\lambda\true,z)$ for clusters of a given input richness, $\lambda\true_{\rm in}$, and redshift, $z_{\rm in}$.  To do so, we compute the observed cluster richness for all halos in the catalogue having a ``true '' richness and redshift equal to the desired input richness/redshift. In practice, we use the criteria $|\lambda\true - \lambda\true_{\rm in}|<0.1\lambda\true_{\rm in}$ and $|z-z_{\rm in}|<0.025$ to select the halos of interest, and replace the assigned richness $\lambda\true$ of
the selected halos by $\lambda\true_{\rm in}$ to avoid introducing artificial scatter in the recovered distribution. Next, we proceed to compute the projected richness of the targeted halos as per Section \ref{sec:method:z_kernel}.  We include percolation effects in this analysis.  If the number of selected clusters in the catalogue is less than $5000$, we generate a new mock catalogue, and iterate the procedure until we arrive at $5000$ independent realizations of $P(\lambda\ob|\lambda\true_{\rm in},z_{\rm in})$.
The key difference between the distribution $P(\lambda\ob|\lambda\true)$ computed here and that obtained in Section~\ref{sec:method:rnd_data} or Section~\ref{sec:method:mock_val} is that the clusters are now correctly embedded within the large scale structure of the Universe, and therefore correlated large scale structure contributes to the incidence of projection effects.

\begin{figure*}
\begin{center}
    \includegraphics[width=0.65 \textwidth]{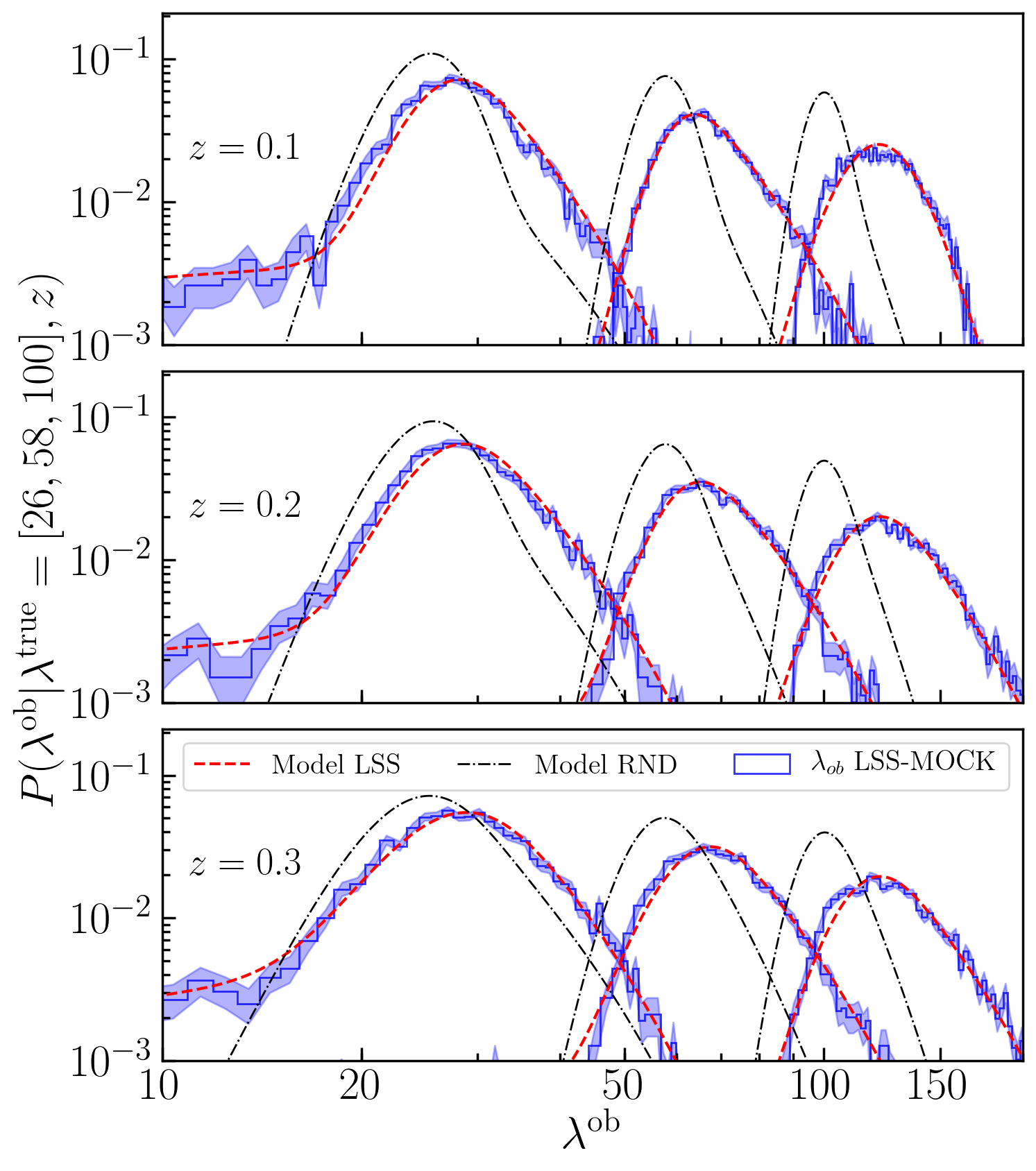}
\end{center}
\caption{$P(\lambda\ob | \lambda\true,z)$ for clusters, including the effect of correlated large-scale structure and percolation, as determined from our simulations. The observed richness, $\lambda\ob$, is computed applying our data calibrated method to include projection effects and background subtraction noise (S. \ref{sec:method:z_kernel}.)  The {\it blue} histogram shows the distributions recovered from the mock data measuring $\lambda\ob$ of $5000$ clusters at their actual position along the large-scale structure for a grid of input redshifts and richnesses (see labels). The shaded area represents the statistical uncertainty of the mock sample.  The {\it red dashed} lines are given by the best-fit model of Equation \ref{eqn:Plobltr_LSS}. For comparison we also include the best-fit model for $P(\lambda\ob | \lambda\true,z)$ obtained in section \ref{sec:method:rnd_data} by injecting clusters at random positions.  Evidently, correlated large scale structure significantly boosts the impact of projection effects.}
\label{fig:Prob_LSS}
\end{figure*}

The resulting distributions in $\lambda\ob$ are shown in Figure \ref{fig:Prob_LSS} for different redshift and richness bins. As expected the magnitude of projection effects increases compared to that obtained when injecting synthetic clusters at random points.  The difference is especially pronounced for the richest clusters --- i.e. the most massive ones --- which live in the most dense environments. Moreover, for low input richnesses the distribution develops a tail toward low $\lambda\ob$ due to percolation: low richness clusters lose member galaxies to richer systems along the line of sight.

Following Equation \ref{eqn:lobNoP}, we model the distributions by setting $\lambda\ob$ to the sum of three random variables,
\begin{equation}
\label{eqn:lob_lss}
\lambda\ob = \lambda\true + \Delta^{\rm bkg} + \Delta^{\rm prj} + \Delta^{\rm prc} .
\end{equation}
Note that in this equation, the term $\Delta^{\rm prj}$ still refers to the noise due to projection effects, but this noise term now incorporates the effects of correlated large-scale structure.  That is, $\Delta^{\rm prj}$ in the above equation is drawn from a different distribution than the noise term $\Delta^{\rm prj}$ appearing in Equation~\ref{eqn:lobNoP}.  In addition, the above equation includes an additional noise term $\Delta^{\rm prc}$, to account for the effect of percolation. We model the distribution of $\Delta^{\rm prc}$ via
\begin{equation}
\label{eqn:P_dprc}
 P(\Delta^{\rm prc}|\lambda\true,z)=(1-f^{\rm msk}) \delta_D(\Delta^{\rm prc}) + \frac{f^{\rm msk}}{\lambda\true} 
 \Theta(-\Delta^{\rm prc})\Theta(\Delta^{\rm prc}+\lambda\true) \, .
\end{equation}
Here $f^{\rm msk}$ represent the fraction of clusters masked by higher-ranked objects. 
The second component of the distribution requires that $\Delta^{\rm prc}$ is equally likely to remove anywhere between no galaxies and
the full $\lambda\true$ galaxies of the projected halo.  That is, $\Delta^{\rm prc} \in [-\lambda\true,0]$.
As for Equation \ref{eqn:P_dprj}, we have found empirically that this distribution provides an accurate fit to our simulated data. 

Combining all three random variables $\Delta^{\rm bkg}$, $\Delta^{\rm prj}$, and $\Delta^{\rm prc}$ we arrive at our final expression for $P(\lambda\ob | \lambda\true,z)$.  We find:
\begin{multline}
\label{eqn:Plobltr_LSS}
P(\lambda\ob | \lambda\true,z) = (1-f^{\rm msk})(1-f^{\rm prj}) \frac{e^{-\frac{(\lambda\ob-\mu)^2}{2\sigma^2}}}{\sqrt{2\pi\sigma^2}} \\
 +\frac{1}{2}\left[ (1-f^{\rm msk})f^{\rm prj} \tau   +  \frac{f^{\rm msk}f^{\rm prj}}{\lambda\true}
    \right]e^{ \frac{\tau}{2}(2\mu +\tau\sigma^2-2\lambda\ob) } {\rm erfc}\left( \frac{\mu +\tau\sigma^2 -\lambda\ob}{\sqrt{2}\sigma} \right)   \\
  + \frac{f^{\rm msk}}{2\lambda\true} \left[ {\rm erfc}\left( \frac{\mu -\lambda\ob-\lambda\true}{\sqrt{2}\sigma} \right)
-  {\rm erfc}\left( \frac{\mu -\lambda\ob}{\sqrt{2}\sigma} \right) \right]  \\ - \frac{f^{\rm msk}f^{\rm prj}}{2\lambda\true} \left[ e^{-\tau \lambda\true} e^{\frac{\tau}{2}(2\mu +\tau\sigma^2-2\lambda\ob) }
 {\rm erfc}\left( \frac{\mu +\tau\sigma^2 -\lambda\ob-\lambda\true}{\sqrt{2}\sigma} \right)   \right] \, .
\end{multline}

The above equation looks complicated, but is conceptually straight forward: projection effects lead to a boost in the richness.  Percolation subtracts out some galaxies because a fraction of the galaxies of low mass halos will have been mistakenly assigned to richer systems.  Finally, there is some Gaussian observational noise on top of these two effects.  Indeed, we view the conceptual simplicity of our model as a key asset, despite the many terms in Equation~\ref{eqn:Plobltr_LSS}.

\begin{figure}
\begin{center}
    \includegraphics[width=0.45 \textwidth]{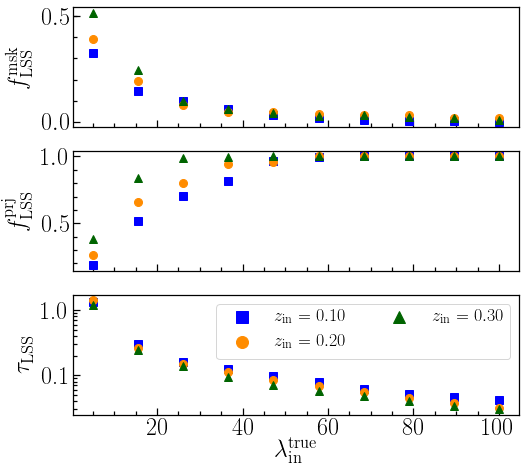}
\end{center}
\caption{Dependence on the input richness and redshift of the model parameters which characterize projection and percolation effects, including the impact of correlated structures along the line of sight (Eq. \ref{eqn:Plobltr_LSS}). The other two model parameters not shown in the figure ($\Delta \mu, \sigma$) are kept fixed to the values shown in Figure \ref{fig:prj_par}.}
\label{fig:prj_par_LSS}
\end{figure}

The best-fit values of the parameters characterizing the impact of correlated structures --- $\tau$, $f^{\rm prj}, f^{\rm msk}$ --- as a function of input richness and redshift are shown in Figure \ref{fig:prj_par_LSS}. As expected, when the impact of correlated structures is included, the fraction of objects affected by projections is larger, reaching one for $\lambda\true \gtrsim 40$. Similarly, the magnitude of the richness perturbations increases --- i.e. $\tau$ values decrease --- compared to the case when only uncorrelated structures are considered (see for comparison Figure \ref{fig:prj_par}). 
Quantitatively, correlated structures boost the richness perturbation $\Delta^{\rm prj}$ by a factor between $2$ and $4$ in the $\lambda\true$ range $20-100$.

The fraction of masked clusters as a function of richness may seem surprising, but this number includes clusters that had even tiny amounts of masking.  The fraction of halos with $\lambda\true=20$ that suffer more than 50\% masking is only 5\%. Moreover, these values are well understood: they {\it must} be there due to purely geometric effects, and their precise value is fairly robust to changes in the details of the percolation.  For instance, changing the \redmapper\ percolation radius by 15\% --- equivalent to assigning the percolation radius of a richness $\lambda = 50$ clusters to a richness $\lambda=100$ cluster --- changes the abundance function by $\approx \pm 2\%$. Such changes have a negligible impact on our cosmological inferences.  



\section{The Impact of Projection Effects on Cosmological Parameter Inference}
\label{sec:cosmo_anal}

To assess the relevance of the proposed modeling on cosmological parameter inference, we perform a cosmological analysis which combines simulated cluster number counts data and weak lensing mass measurements. Specifically, we reproduce the cosmological analysis performed in a companion paper (Costanzi et al, in preparation), where we use the model developed here to place cosmological constraints using the SDSS \redmapper\ catalogue.  A detailed description of the full likelihood is presented in that work.  In addition, a similar analysis using the DES Y1 data is forthcoming. Here, we limit ourselves to a brief summary.

Our synthetic data vector is derived from the halo catalogue introduced in Section~\ref{sec:method:z_kernel} implementing our procedure to include projection effects in simulated data.
The data vector consists of the number of galaxy clusters
in five richness bins --- the bin edges are $\Delta\lambda\ob = [20,27.9,37.6,50.3,69.3,140]$ --- and two redshift bins --- $z\in[0.1,0.2)$ and $z\in [0.2,0.3)$. We assume a weak lensing analysis enables us to recover the mean mass for the clusters in each richness/redshift bin.  These mean masses are taken by computing the mean halo mass of the clusters in a bin. Random noise is added by using a multivariate Gaussian distribution defined by the covariance matrix for the weak lensing mass estimates.  The latter is computed starting from the weak lensing mass errors derived for the \redmapper\ SDSS catalogue by \citet{Simet2016}, and assuming that errors associated with the multiplicative shear bias, photometric redshift uncertainties, projection effects and triaxiality, are all perfectly correlated across all richness bins.   The statistical shape noise is taken directly from the weak lensing analysis by \citet{Simet2016}. We do not rescale the shape noise errors to account for the number of clusters in the simulations and each bin: the errors are exactly those from \citet{Simet2016}.  The uncertainty in the amplitude of the mass--richness relation for the simulated data vector is $\approx 4.5\%$, smaller than the corresponding error budget in SDSS (the simulation has a larger number of clusters, and we have two redshift bins, each with errors identical to the single-redshift bin result for SDSS), but comparable to the error recently reported by \citet{desy1wl} with the DES Y1 cluster sample.

Given these synthetic data vectors and their covariance matrix, we sample the appropriate likelihood distribution using the {\ttfamily emcee} package\footnote{http://dan.iel.fm/emcee/} \citep{Foreman2013} to explore the parameter space.  The likelihood is modeled as a Gaussian distribution.  The expectation value for the number counts is computed by integrating Equation \ref{eqn:dndz} over the relevant richness and redshift bin. Similarly, the expected mean cluster mass is computed by weighting Equation \ref{eqn:dndz} by the halo mass and integrating over the $\lambda\ob$ and $z$ bins. The covariance matrix for the abundances includes both Poisson noise and sample variance \cite[e.g.,][]{hukravtsov03}, while the covariance matrix for the weak lensing mass estimates is described above.  We assume no covariance between the weak lensing mass data and the abundance data.

To assess the relevance of projection effects on parameter inference we consider four models for the scatter between the {\it true} and {\it observed} richness:
\begin{enumerate}
	\item{ We account for projection effects using Equation \ref{eqn:Plobltr_LSS} with the best-fit values recovered from the analysis as our model parameters.}
	\item{We neglect the effect of correlated structures by using $P(\lambda\ob|\lambda\true)$ calibrated from cluster placed at random positions.}
    \item{We neglect the effect of masking setting $f^{\rm msk}$ to zero in Equation \ref{eqn:Plobltr_LSS}.}
    \item{We ignore both masking and projection effects by setting $\lambda\ob = \lambda\true+ \Delta^{\rm bkg}$, i.e. we consider only the Gaussian observational noise term. This model is typical of analyses to date, and ignore the impact of projection effects on the shape of $P(\lambda\ob|\lambda\true)$ \citep[e.g.][]{Rozo2007,Rozo2010}.}
\end{enumerate}
In all cases, we simultaneously constrain the cosmological parameters $\sigma_8$ and $\Omega_m$, and the richness-mass relation parameters $\ln\lambda_0$, $\alpha$, and $\sigma_{\rm intr}$ (see Equations \ref{eqn:OMR} and \ref{eqn:sig_lm}).

\begin{figure*}
\begin{center}
    \includegraphics[width=\textwidth]{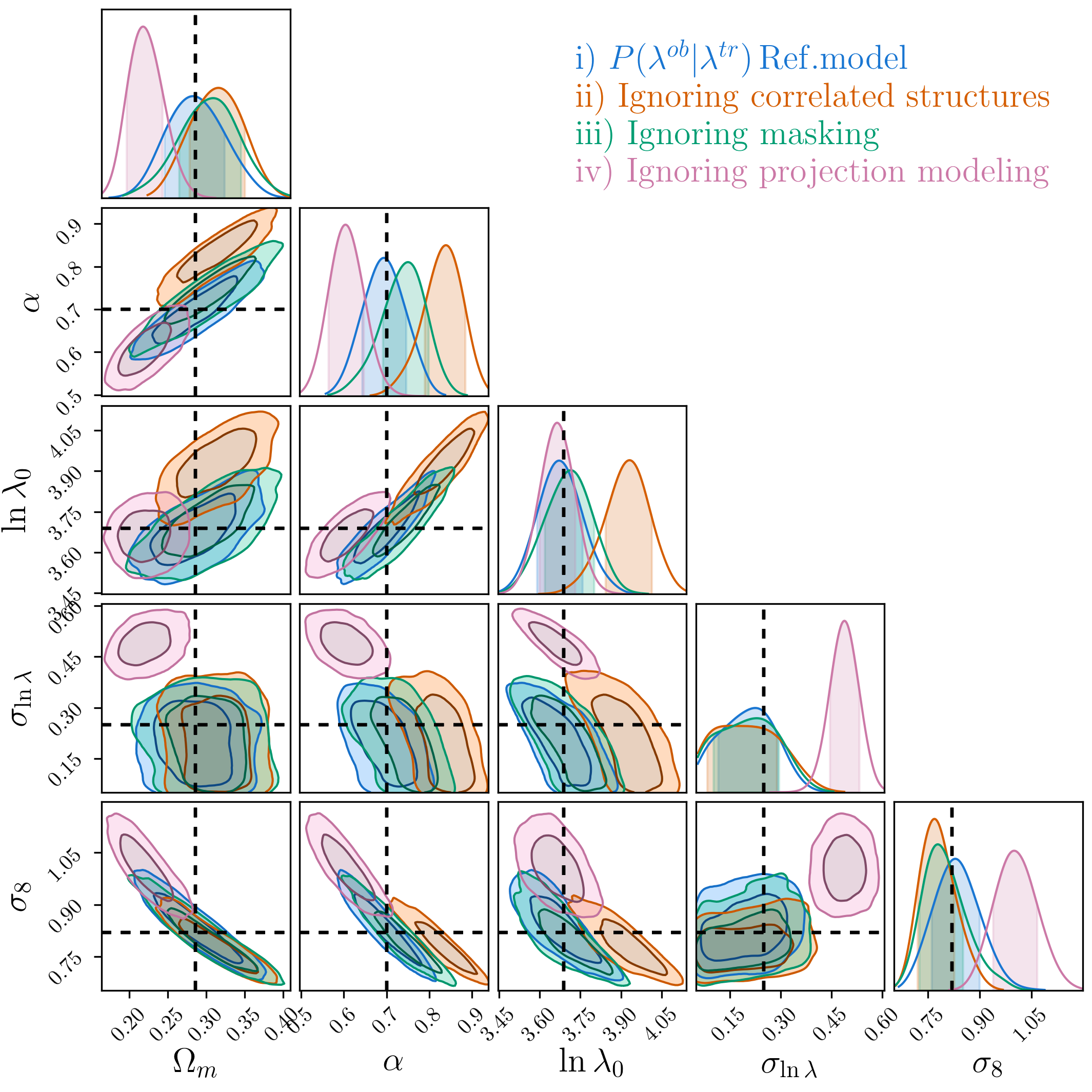}
\end{center}
\caption{Posterior on cosmological and richness--mass relation parameters for our synthetic data set. The synthetic data was generated by applying our procedure to include projection effects and observational noise on a simulated halo catalog as detailed in Section \ref{sec:method:z_kernel}. We considered four separate cases.  The {\it blue} contours are obtained by accounting for projection effects as advocated in this work. The {\it orange} contours are derived using $P(\lambda\ob|\lambda\true,z)$ as calibrated from clusters injected at random positions , i.e. neglecting the effect of correlated structures.
Green contours are obtained neglecting masking effects, that is, setting the parameter $f^{\rm msk}$ to zero in Equation \ref{eqn:Plobltr_LSS}. Finally, the {\it pink} contours are derived ignoring both projection and masking effects by setting $P(\lambda\ob|\lambda\true,z)$ equal to the Gaussian noise characterizing observational noise.  We can see that failing to properly model projection effects can potentially introduces large biases in the inferred cosmological parameters.
}
\label{fig:TriPlot}
\end{figure*}

The results of our analyses are shown in Figure \ref{fig:TriPlot}.
As expected, model $(i)$ recovers both the cosmological and richness--mass relation input parameters. When neglecting the effects of correlated structures or masking -- model $(ii)$ and $(iii)$ -- the input cosmological parameters are still recovered within errors, though for model $(ii)$ there is a $\sim 1 \sigma$ bias in the recovered cosmological parameter. The relatively small bias reflects the fact that in model $(ii)$ the smaller skewness of $P(\lambda\ob|\lambda\true)$ is compensated by a steeper slope and larger normalization of the richness-mass relation. Turning to model $(iii)$, masking effects are rare, and impact primarily low richness objects -- $\lambda \lesssim 20$ --   so we are able to correctly recover the fiducial richness-mass relation parameters despite ignoring percolation effects. By contrast, for model $(iv)$ the recovered richness--mass relation is biased, and the corresponding cosmological constraints disfavor the input cosmology at more than $2\sigma$.  Unsurprisingly, the recovered scatter $\sigma_{\rm intr} \sim 0.5$ is significantly larger than the input scatter for the simulation, $\sigma_{\rm intr}=0.25$.
Despite this extra scatter absorbed by $\sigma_{\rm intr}$, the mismatch between the true \emph{shape} of $P(\lambda\ob|M,z)$ and the model assumed in $(iv)$ results in a biased cosmological inference. 

While this analysis clearly shows the importance of accounting for projection effects, it is important to stress that the level of biases caused by an incorrect calibration of $P(\lambda\ob|\lambda\true)$ depends on the size of the cluster catalogue, the accuracy of the mass calibration, and the flexibility of the richness-mass relation adopted. In particular for a mock catalogue having the same statistical properties as those of the SDSS \redmapper\ catalog and its associated weak lensing data, we find that the cosmological parameter posteriors are only minimally biased.  That is, the modeling described in this paper is necessary for Stage III dark energy experiments, but not for stage II.

Finally we test the robustness of our results to the details of the procedure used to calibrate the impact of projection effects, specifically the input cosmology and richness--mass relation parameters of the simulation, and the percentile used to calibrate the width $\sigma_z(z)$ of the projection kernel $w(\Delta z,z)$.
To this end, following Section \ref{sec:prj_LSS}, we re-calibrate $P(\lambda\ob|\lambda\true,z)$ three times as follows: $i)$ using mock catalogues generated by perturbing the richness-mass relation parameters used to populate the simulation by one standard deviation, $ii)$ approximating the lower envelope of the redshift kernel with the $90$ percentile (see \S \ref{sec:method:z_kernel} for details) and $iii)$ assuming four different input cosmologies falling outside the $95\%$ confidence region of our fiducial posterior distribution. As for the latter, we use Equation \ref{eqn:f_tau} to derive the cosmology dependent shift of the projection effect parameters used in Equation \ref{eqn:Plobltr_LSS}. For each new calibration of $P(\lambda\ob|\lambda\true,z)$ we repeat the cosmological analysis keeping the same fiducial mock data and covariance matrices described before. That is, the input data vector into our likelihood is always the same.

\begin{figure*}
\begin{center}
    \includegraphics[width= \textwidth]{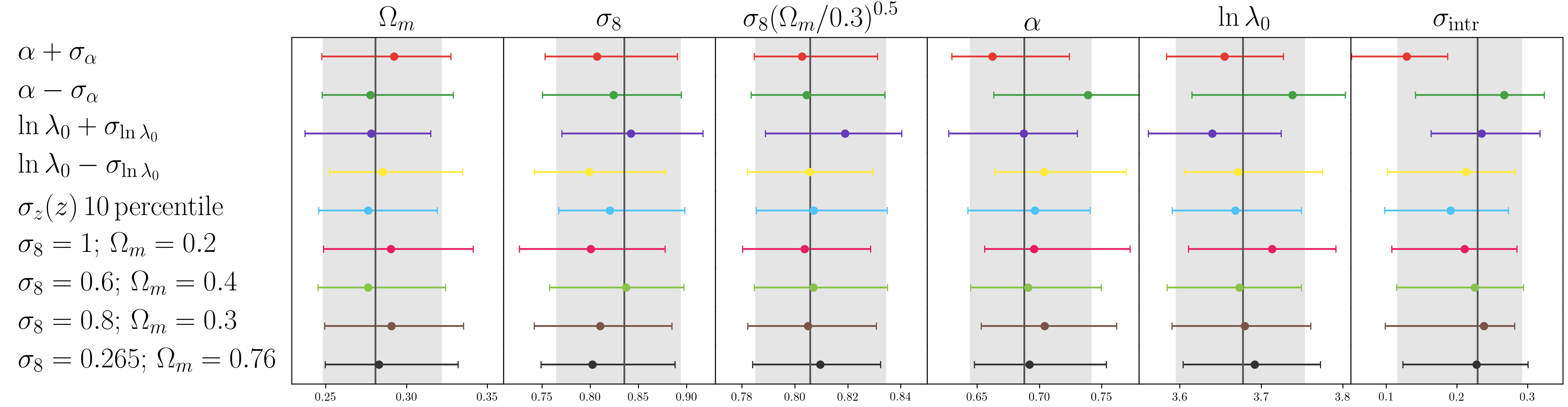}
\end{center}
\caption{Comparison of the $68\%$ confidence regions derived from our synthetic data set to the posteriors from our fiducial analysis after perturbing the input model parameters relevant for the calibration of $P(\lambda\ob|\lambda\true,z)$. The shaded area correspond to $68\%$ confidence region derived using our reference model. The colored error bars corresponds to the results derived using $P(\lambda\ob|\lambda\true,z)$ calibrated varying: the richness-mass relation parameters ($i-iv$), the percentile used to define the redshift kernel ($v$) or the input cosmology ($vi-ix$) (see labels in the plot).}
\label{fig:syst}
\end{figure*}

The results of these analyses are summarized in Figure \ref{fig:syst} which compares the $68\%$ confidence regions derived using the different calibrations of $P(\lambda\ob|\lambda\true,z)$ (colored error bars) with the results obtained using the reference model ({\it shaded gray} area).
We consistently recover the correct cosmological and richness-mass relation parameters regardless of the details of how we calibrate our projection effects model.  In short, our calibration procedure is both robust, and sufficient for enabling accurate cosmological parameter estimates from current cluster surveys.

%


\section{Summary and Conclusion}
\label{sec:conc}

We have developed a new algorithm for quantitatively characterizing projection effects in cluster catalogues, and applied this algorithm to the \redmapper\ SDSS DR8 cluster catalogue.  Our method combines real data with $N$-body simulations to correctly account for the effects of correlated large-scale structure on projection effects.
Specifically, we use the real data to calibrate the observational noise in richness estimates due to photometric noise and background subtraction, and to validate our method for incorporating projection effects into simulations. By comparing the probability distributions for $\lambda\ob$ given $\lambda\true$ as recovered in both simulated and real data we are able to demonstrate the validity of our method for adding projection effects in simulations.  Finally, we use these quantitatively validated numerical simulations to characterize the impact of correlated large-scale structures in the recovered richness measurements. 

We find that projection effects can substantially alter the observed richness, and that the effect is especially strong in rich galaxy clusters due to the abundance of correlated structures around these systems.  By performing a cosmological analysis of a synthetic data set with a known underlying cosmology, we demonstrate that explicitly modeling projection effects is necessary in order to derive unbiased cosmological constraints from upcoming photometrically selected cluster catalogues.

The fully calibrated distribution $P(\lambda\ob|\lambda\true,z)$ recovered in this analysis will be utilized in a companion paper (Costanzi et al, in preparation) to derive cosmological constraints from the SDSS \redmapper\ cluster sample.  This formalism will further be applied to the analysis of the DES Year 1 \redmapper\ cluster catalogue.
Following the tests performed in Section \ref{sec:cosmo_anal}, we will assess the robustness of the inferred results to the $P(\lambda\ob|\lambda\true,z)$ calibration by repeating the cosmological analysis with different calibrations of the projection effects parameters, as obtained using a variety of cosmologies and input richness--mass relations.
As demonstrated in Section~\ref{sec:cosmo_anal}, similar analyses to that carried out here will be necessary for all future photometric cluster surveys seeking to place cosmological constraints.

We again emphasize that the calibration of projection effects in this work is specific to the SDSS \redmapper\ cluster sample, and that other cluster finding algorithms and/or other data sets will require a full recalibration of projections effects, as these depend both on the algorithm employed, and the photometric properties of the survey.  Ultimately, a full calibration of $P(\lambda\ob|\lambda\true,z)$ should describe the full complexity of the mapping between halos and clusters, including e.g. miscentering, triaxiality, blending and fragmentation effects, tested on appropriately tuned synthetic galaxy catalogues from cosmological simulations. Moreover, additional work is required to enable forward modeling of projection effects for cluster lensing.  In the mean time, cosmological analyses must rely on ``backwards modeling'' of projection effects for these two probes \citep[e.g.][]{Simet2016,desy1wl,baxteretal16}.  Nevertheless, the success of the model advocated for in this work is an important step forward in the modeling of photometric cluster samples, and provides a critical stepping stone in our quest to saturate the statistical limit of ongoing and future photometric cluster surveys.


\section*{Acknowledgements}
\label{sec:acknowledgements}

This paper has gone through internal review by the DES collaboration.
ER is supported by DOE grant DE-SC0015975 and by the Sloan Foundation, grant FG-2016-6443. YP is
supported by DOE grant DE-SC0015975. 
Support for DG was provided by NASA through Einstein Postdoctoral
Fellowship grant number PF5-160138 awarded by the Chandra X-ray
Center, which is operated by the Smithsonian Astrophysical Observatory
for NASA under contract NAS8-03060.  RM is supported by the Department of Energy Cosmic Frontier program, grant DE-SC0010118.
Funding for the DES Projects has been provided by the U.S. Department of Energy, the U.S. National Science Foundation, the Ministry of Science and Education of Spain, 
the Science and Technology Facilities Council of the United Kingdom, the Higher Education Funding Council for England, the National Center for Supercomputing 
Applications at the University of Illinois at Urbana-Champaign, the Kavli Institute of Cosmological Physics at the University of Chicago, 
the Center for Cosmology and Astro-Particle Physics at the Ohio State University,
the Mitchell Institute for Fundamental Physics and Astronomy at Texas A\&M University, Financiadora de Estudos e Projetos, 
Funda{\c c}{\~a}o Carlos Chagas Filho de Amparo {\`a} Pesquisa do Estado do Rio de Janeiro, Conselho Nacional de Desenvolvimento Cient{\'i}fico e Tecnol{\'o}gico and 
the Minist{\'e}rio da Ci{\^e}ncia, Tecnologia e Inova{\c c}{\~a}o, the Deutsche Forschungsgemeinschaft and the Collaborating Institutions in the Dark Energy Survey. 

The Collaborating Institutions are Argonne National Laboratory, the University of California at Santa Cruz, the University of Cambridge, Centro de Investigaciones Energ{\'e}ticas, 
Medioambientales y Tecnol{\'o}gicas-Madrid, the University of Chicago, University College London, the DES-Brazil Consortium, the University of Edinburgh, 
the Eidgen{\"o}ssische Technische Hochschule (ETH) Z{\"u}rich, 
Fermi National Accelerator Laboratory, the University of Illinois at Urbana-Champaign, the Institut de Ci{\`e}ncies de l'Espai (IEEC/CSIC), 
the Institut de F{\'i}sica d'Altes Energies, Lawrence Berkeley National Laboratory, the Ludwig-Maximilians Universit{\"a}t M{\"u}nchen and the associated Excellence Cluster Universe, 
the University of Michigan, the National Optical Astronomy Observatory, the University of Nottingham, The Ohio State University, the University of Pennsylvania, the University of Portsmouth, 
SLAC National Accelerator Laboratory, Stanford University, the University of Sussex, Texas A\&M University, and the OzDES Membership Consortium.

Based in part on observations at Cerro Tololo Inter-American Observatory, National Optical Astronomy Observatory, which is operated by the Association of 
Universities for Research in Astronomy (AURA) under a cooperative agreement with the National Science Foundation.

The DES data management system is supported by the National Science Foundation under Grant Numbers AST-1138766 and AST-1536171.
The DES participants from Spanish institutions are partially supported by MINECO under grants AYA2015-71825, ESP2015-66861, FPA2015-68048, SEV-2016-0588, SEV-2016-0597, and MDM-2015-0509, 
some of which include ERDF funds from the European Union. IFAE is partially funded by the CERCA program of the Generalitat de Catalunya.
Research leading to these results has received funding from the European Research
Council under the European Union's Seventh Framework Program (FP7/2007-2013) including ERC grant agreements 240672, 291329, and 306478.
We  acknowledge support from the Australian Research Council Centre of Excellence for All-sky Astrophysics (CAASTRO), through project number CE110001020.

We thank Peikai Li for providing an analytical solution of the mean fraction of overlapping area (equation \ref{eqn:f_over})  

This manuscript has been authored by Fermi Research Alliance, LLC under Contract No. DE-AC02-07CH11359 with the U.S. Department of Energy, Office of Science, Office of High Energy Physics. The United States Government retains and the publisher, by accepting the article for publication, acknowledges that the United States Government retains a non-exclusive, paid-up, irrevocable, world-wide license to publish or reproduce the published form of this manuscript, or allow others to do so, for United States Government purposes.


\appendix
\section{Analytic derivation of projection effect parameters}
\label{app:prj_eff}

Here we present a model to estimate analytically the projection effect parameters for a given input cosmology and richness-mass relation.
The results of this section are used in the main analysis to estimate the response of the projection effect parameters to a shift of the cosmological parameters $\sigma_8$ and $\Omega_m$. 

The number of clusters expected to fall inside the light cone defined by the angular size of a cluster having radius $R(\lambda\true)$ at redshift $z\true$  is given by:
\begin{multline}
\bar{N}_{\Omega} (\lambda\true,z\true) =  \int_0^{\infty} \de z\ \frac{\de V}{\de z \de \Omega} 
\int_0^{\infty} \de \lambda\ \Omega(\lambda\true,z\true,\lambda,z) \\ \int_0^{\infty} \de M\ n(M,z) P(\lambda | M,z) \, ,
\end{multline}
where the angular aperture within which we count objects is defined by the sum of the angular radii, $\vartheta$, of the objects considered: 
\begin{eqnarray}
\vartheta(\lambda,z)&=&\frac{R(\lambda)}{D_A(z)} \\
\Omega(\lambda\true,z\true,\lambda,z)&=&2\pi \left[1-\cos \left( \vartheta(\lambda\true,z\true)+\vartheta(\lambda,z) \right) \right] \, , \nonumber
\end{eqnarray}

For a cluster placed at a random position, the expected shift on the observed richness due to projection effects, $\bar{\Delta}^{\rm prj}$, can be estimated by weighing the previous equation by the number of member galaxies that each cluster has inside the appropriate light cone, as these are the galaxies that can be ``shared'' with the main halo at $z\true$. I.e., following Equation \ref{eqn:prj}:
\begin{multline}
\label{eqn:barD}
\bar{\Delta}^{\rm prj} (\lambda\true,z\true) =  \int_0^{\infty} \de z\ \frac{\de V}{\de z \de \Omega}  
\int_0^{\infty} \de \lambda\  \Omega(\lambda\true,z\true,\lambda,z) \\ \Delta^{\rm prj} (\lambda,z)  \int_0^{\infty} \de M\ n(M,z) P(\lambda | M,z) \, ,
\end{multline}
where we have defined
\begin{equation}
\Delta^{\rm prj} (\lambda,z)=w(z- z\true,\sigma_z(z)) \bar{f}(\lambda\true,z\true,\lambda,z)\lambda \, ,
\end{equation}
$w(z- z\true,\sigma_z(z))$ is the redshift weight defined in Equation \ref{eqn:w_z} and $\bar{f}(\lambda\true,z\true,\lambda,z)$ is the mean fraction of overlapping area of halos inside the appropriate light cone, i.e. in the angular aperture $\Omega$ defined by $\vartheta(\lambda\true,z\true)$ and $\vartheta(\lambda,z)$. 
Using the flat-sky approximation the integral over the solid angle $\Omega$ can be solved analytically, and the mean fraction of overlapping area reads:
\begin{equation}
\label{eqn:f_over}
 \bar{f}(\lambda\true,z\true,\lambda,z) = \left(1+\frac{\vartheta(\lambda,z)}{\vartheta(\lambda\true,z\true)} \right)^{-2}
\end{equation}

Assuming $N_{\Omega} (\lambda\true,z\true)$ to follow a Poisson distribution, the variance of $\Delta^{\rm prj}$ can be computed as:
\begin{multline}
\label{eqn:varD}
{\rm Var}(\Delta^{\rm prj} (\lambda\true,z\true)) =  \int_0^{\infty} \de z\ \frac{\de V}{\de z\ \de\ \Omega} \int_0^{\infty} \de\ \lambda \,  \Omega(\lambda\true,z\true,\lambda,z) \\ \left(\Delta^{\rm prj} (\lambda,z)\right)^2  \int_0^{\infty} \de M\ n(M,z) P(\lambda | M,z) \, .
\end{multline}

Assuming $P(\Delta^{\rm prj})$ to follow Equation \ref{eqn:P_dprj}, we can relate the latter two derived quantities to the two model parameters $f^{\rm prj}$ and $\tau$. In particular, from Equation \ref{eqn:P_dprj} it follows that:

\begin{eqnarray}
\label{eqn:mean_var}
\bar{\Delta}^{\rm prj} &=& \frac{f^{\rm prj}}{ \tau} \\ 
{\rm Var}(\Delta^{\rm prj}) &=& \frac{f^{\rm prj}(2-f^{\rm prj} )}{\tau^2} \, . \nonumber
\end{eqnarray}
Solving for $f^{\rm prj}$ and $\tau$ one gets:
\begin{eqnarray}
\label{eqn:f_tau}
f^{\rm prj} &=& 2\frac{(\bar{\Delta}^{\rm prj})^2}{(\bar{\Delta}^{\rm prj})^2+{\rm Var}(\Delta^{\rm prj}) } \\ 
\tau &=& 2\frac{\bar{\Delta}^{\rm prj}}{(\bar{\Delta}^{\rm prj})^2+{\rm Var}(\Delta^{\rm prj})} \nonumber
\end{eqnarray}

To verify the above assumptions we make use of mock catalogues. Specifically, from the analysis performed in Section \ref{sec:method:mock_val} we compute $\bar{\Delta}^{\rm prj}$ and ${\rm Var}(\Delta^{\rm prj})$ for the various input $(\lambda_{in},z_{in})$ tested, and compare these values with the ones derived from Equations \ref{eqn:barD} and \ref{eqn:varD}. 
\begin{figure}
\begin{center}
    \includegraphics[width=0.50 \textwidth]{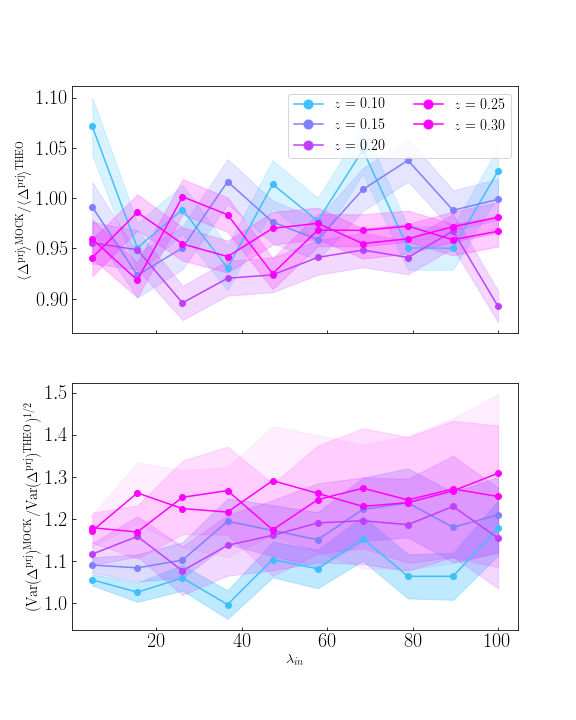}
\end{center}
\caption{Comparison of the theoretical predictions of $\bar{\Delta}^{\rm prj}$ ({\it upper panel}) and ${\rm Var}(\Delta^{\rm prj})$ ({\it lower} panel) with the values recovered from the {\it random} mock catalogue for different input richnesses and redshifts. The bands correspond to the errors on the estimate of the mean and variance from the mock catalogue.}
\label{fig:meanD}
\end{figure}
The result of this test is shown in Figure \ref{fig:meanD} for different richness and redshift bins. Our model for the expectation value of $\Delta^{\rm prj}$ is only slightly biased toward higher values, whereas our theoretical predictions for the variance are biased low by $\sim 10\%-20\%$. The main reason for the latter disagreement is due to correlated structures along the line of sight in the mock catalogue: even if we are considering $\Delta^{\rm prj}$ for clusters placed at random positions, the objects which appear to be in projection are not randomly distributed along the radial direction. We have verified this explicitly: when we repeat the analysis performed in Section \ref{sec:method:mock_val} after randomizing the angular positions of all clusters in the mock catalogue, then the simulations agree with our theoretical model at the 5\% level.

\begin{figure}
\begin{center}
    \includegraphics[width=0.50 \textwidth]{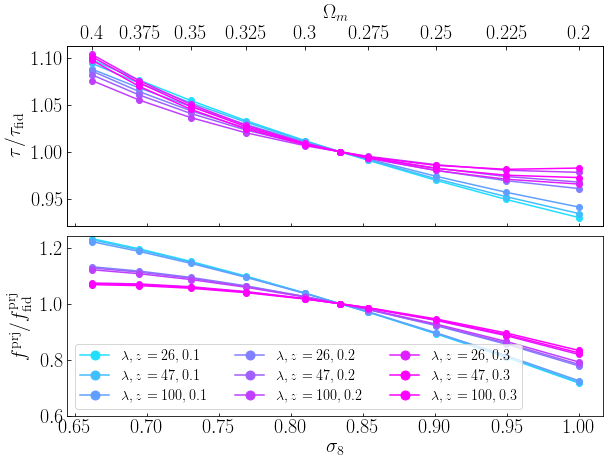}
\end{center}
\caption{Predicted shift of $\tau$ and $f^{\rm prj}$ relative to their fiducial cosmology values for different combinations of $(\sigma_8,\Omega_m)$ satisfying the relation $\sigma_8 (\Omega_m/0.3)^{0.58}=0.80$. Different colors correspond to different input richnesses and redshifts (see figure labels).}
\label{fig:tau_f_s8}
\end{figure}

Even if the theoretical model proposed is not capable of fully reproduce the mock results, we can use it to estimate the impact of cosmology on projection effects. In particular, we exploit the analytic model to evaluate the relative shift of $\tau$ and $f^{\rm prj}$ as a function of cosmology. As an example to illustrate the magnitude of the variation of $\tau$ and $f^{\rm prj}$, we consider a set of cosmologies satisfying the relation: $\sigma_8 (\Omega_m/0.3)^{0.58}=0.80$. The latter corresponds to the degeneracy direction between $\sigma_8$ and $\Omega_m$ recovered from the mock cosmological analysis of Section \ref{sec:cosmo_anal}. Figure \ref{fig:tau_f_s8} shows the shift of $\tau$ and $f^{\rm prj}$ relative to their values on the fiducial cosmology ($\sigma_8=0.82$, $\Omega_m=0.286$), as a function of the cosmological model for different input richnesses and redshifts. The magnitude of the shift is almost independent on the input richness, and shows some modest variation with redshift.  
The changes in $P(\lambda\ob|\lambda\true,z)$ shown in Figure \ref{fig:PDFvsCosmo} correspond to the shifts in $\tau$ and $f^{\rm prj}$ computed via Equation \ref{eqn:f_tau} for $30$ pairs of ($\sigma_8,\Omega_m$) values randomly sampled from the posterior derived from the cosmological analysis detailed in section \ref{sec:cosmo_anal}.

The above analytical derivation can be ideally extended to real clusters accounting for correlated structures.
Perhaps the simplest way to account for them is to boost the expected number of halos around the clusters according to the halo-halo correlation function, i.e. multiplying the halo mass function in Equation \ref{eqn:barD} and \ref{eqn:varD} by
\begin{equation}
(1 + b(M,z) b(\langle M (\lambda\true)\rangle,z\true) \xi_{\rm NL}(|r(z)-r(z\true)|,\bar{z})) \, ,
\end{equation}
where $b(M,z)$ represent the halo bias and $\xi_{\rm NL}(|r(z)-r(z\true)|,\bar{z}))$ is the non-linear matter correlation function at the mean redshift $\bar{z}=(z+z\true)/2$ and co-moving distance $|r(z)-r(z\true)|$.
To account for exclusion effects we include the condition:
\begin{equation*}
 b(M,z) b(\langle M (\lambda\true)\rangle,z\true) \xi_{\rm NL}(|r(z)-r(z\true)|,\bar{z}))=0  
\end{equation*}
if $|r(z)-r(z\true)| < R(\lambda\true)(1+z\true)$.

As before we compare our analytical derivations with results from the mock data, this time including the effect of correlated structures (S. \ref{sec:prj_LSS}). In this case both predictions for $\bar{\Delta}^{\rm prj}$ and ${\rm Var}(\Delta^{\rm prj})$ underestimate the mock data results by more than $50\%$. Note that a 20\% difference between our model and numerical simulations was already observed when considering the impact of projection effects about random points, a difference we attributed to clustering of large scale structure.  It is therefore not surprising that a further underestimate of projection effects occurs when considering the impact of projection effects about dark matter halos.  In particular, note that our model explicitly ignores the contribution of higher order (e.g. 3-point and 4-point) clustering in our estimate.  We postpone a detailed analytic model of such effects to future work.



\bibliographystyle{mnras}
\bibliography{bibliography.bib} 


\bsp
\label{lastpage}
\end{document}